\documentclass{article}
\usepackage{amsmath}
\usepackage{amssymb}
\usepackage{mathrsfs}
\usepackage{tikz}
\usepackage[top=1in,bottom=1in,left=1in,right=1in]{geometry}
\usepackage{url}
\usepackage[hidelinks]{hyperref}
\usepackage{array}
\usepackage{multirow}

\usepackage{ulem}

\newcommand{\p}[2]{p_{#1}^{\text{#2}}}
\newcommand{\Obs}[2]{\widehat{O}_{#1}^{\text{#2}}}
\newcommand{\db}{\bar\delta}
\newcommand{\del}[1]{\delta^{\text{#1}}}
\newcommand{\delb}[1]{\bar\delta^{\text{#1}}}
\newcommand{\dal}{\Delta\alpha}
\newcommand{\dalhad}{\Delta\alpha^{(5)}_{\text{had}}}
\newcommand{\alsmz}{\alpha_s(m_Z)}
\newcommand{\almz}{\alpha(m_Z)}
\newcommand{\sineff}[1]{\sin^2\theta_{\text{eff}}^{#1}}
\newcommand{\inv}{_{\text{inv}}}
\newcommand{\had}{_{\text{had}}}
\newcommand{\AFB}[1]{A_{\text{FB}}^{#1}}
\newcommand{\Asy}[1]{\mathcal{A}_{#1}}
\newcommand{\flag}[1]{\texttt{#1}}
\newcommand{\Op}{\mathscr{O}}
\newcommand{\Ord}[1]{\mathcal{O}\left(#1\right)}

\newcommand{\tree}[1]{{\left[#1\right]}^{\text{\tiny{(0)}}}}

\newcommand{\pizz}{\pi_{zz}}
\newcommand{\piww}{\pi_{ww}}
\newcommand{\piwwo}{\pi_{ww}^0}
\newcommand{\piggp}{\pi_{\gamma\gamma}^{\prime}}
\newcommand{\pigz}{\pi_{\gamma z}}
\newcommand{\pizzp}{\pi_{zz}^{\prime}}

\newcommand{\be}{\begin{equation}}
\newcommand{\bea}{\begin{eqnarray}}
\newcommand{\beq}[1]{\begin{equation}\label{#1}}
\newcommand{\beqa}[1]{\begin{eqnarray}\label{#1}}
\newcommand{\eq}[1]{Eq.~(\ref{#1})}
\newcommand{\eqs}[2]{Eqs.~(\ref{#1}) and~(\ref{#2})}

\newcommand{\ee}{\end{equation}}
\newcommand{\eea}{\end{eqnarray}}
\newcommand{\eeq}{\end{equation}}
\newcommand{\eeqa}{\end{eqnarray}}

\begin{document}

\begin{titlepage}
\noindent
\begin{flushright}
{\small  ~} \\  
\end{flushright}

\vspace{0.5cm}

\begin{center}
  \begin{Large}
    \begin{bf}
Precision Electroweak Analysis after the Higgs Boson Discovery
   \end{bf}
  \end{Large}
\end{center}
\vspace{0.2cm}
\begin{center}
\begin{large}
James D. Wells and Zhengkang Zhang \\
\end{large}
  \vspace{0.3cm}
  \begin{it}
Michigan Center for Theoretical Physics (MCTP) \\
Department of Physics, University of Michigan, Ann Arbor, MI 48109
\end{it}

\end{center}

\center{\today}

\begin{abstract}

Until recently precision electroweak computations were fundamentally uncertain due to lack of knowledge about the existence of the Standard Model Higgs boson and its mass. For this reason substantial calculational machinery had to be carried along for each calculation that changed the Higgs boson mass and other parameters of the Standard Model. Now that the Higgs boson is discovered and its mass is known to within a percent, we are able to compute reliable semi-analytic expansions of electroweak observables. We present results of those computations in the form of expansion formulae. In addition to the convenience of having these expressions, we show how the approach makes investigating new physics contributions to precision electroweak observables much easier.

\end{abstract}

\tableofcontents
\vspace{1cm}

\end{titlepage}

\setcounter{page}{2}

\section{Introduction}

Precision electroweak analysis has played an important role in testing the Standard Model (SM) and constraining new physics. Now this program has entered a new era with  the discovery of the Higgs boson~\cite{Aad:2012tfa,Chatrchyan:2012ufa}. On one hand, the sub-percentage-level determination of the Higgs boson mass~\cite{Aad:2012tfa,Chatrchyan:2012ufa,Chatrchyan:2012jja} constitutes the last piece of a complete set of input observables. Electroweak observables can now be calculated to unprecedented accuracy, leading to unprecedented  sensitivity to new physics beyond the SM. On the other hand, measurements of the Higgs observables, such as its decay widths and branching ratios, will push our understanding of elementary particle physics to more stringent tests. In this paper we focus on the former aspect. For the latter aspect, see e.g.~\cite{Almeida:2013jfa}.

The standard approach of precision electroweak analysis is to perform a $\chi^2$ analysis, which involves varying the model parameters, or equivalently, a set of input observables to minimize the $\chi^2$ function. In practice, this can be facilitated by an expansion about some reference values of the input, since we have a set of well-measured input observables that allows little variation. We present such an expansion formalism, and apply it to deriving constraints on new physics models. Most of the numerical results in this paper reflect state-of-the-art calculations of the electroweak observables, as implemented in the ZFITTER package~\cite{Bardin:1999yd,Arbuzov:2005ma}.

Our paper is organized as follows. We first review the definition of the electroweak observables under consideration in Section~\ref{sec:SMpO}. Then in Section~\ref{sec:formalism} we present the expansion formalism for calculating the SM and new physics contributions to the observables. The result will be that given the values of 6 input observables, and the new physics model, all observables can be easily calculated. The tools needed in this calculation, including the reference values of all observables, and the expansion coefficients, are presented. Next, we illustrate how to use the formalism by working out some new physics examples in Section~\ref{sec:NPex}. Finally, in Section~\ref{sec:summary} we summarize.

\section{Standard Model parameters and observables}\label{sec:SMpO}

The parameters of the SM include the gauge couplings $g_3$, $g_2$, $g_1$, the Yukawa couplings $y_f$, flavor angles, the Higgs vacuum expectation value $v$ and self-coupling $\lambda$. For the purpose of precision electroweak analysis, with inconsequential errors we can treat all Yukawa couplings except that for the top quark as constants, and correspondingly set the lepton and light quark masses to their default values in ZFITTER (see~\cite{Bardin:1999yd}). Then there are six parameters\footnote{We do not include flavor CKM angles in our calculations since all standard precision electroweak observables do not substantively depend on these angles.} in the theory:
\beq{SMparam}
\{ g_3,\; g_2,\; g_1,\; y_t,\; v,\; \lambda \}.
\eeq

There are an infinite number of SM observables that can be defined. They correspond to well-defined quantities that are measured in experiments. The SM predicts each observable as a function of the parameters in \eq{SMparam}. The success of the SM relies on the fact that the prediction for all observables agree with precision measurements, with suitable choices of the parameters. If some new physics beyond the SM were to exist, it could potentially destroy the agreement. Thus, precision analysis enables us to put stringent constraints on new physics models. In this paper we focus on the following list of observables, mostly relevant to precision tests of the electroweak theory.
\begin{itemize}
  \item Pole mass of the particles: $m_Z$, $m_W$, $m_t$, $m_H$.
  \item Observables associated with the strengths of the strong, weak, and electromagnetic interactions: $\alsmz$, $G_F$, and $\almz$. The Fermi constant $G_F$ is defined via the muon lifetime~\cite{Beringer:1900zz}. $\almz$ is related to the fine structure constant $\alpha_0$ defined in the Thomson limit via
      \be
      \almz = \frac{\alpha_0}{1-\dal_\ell-\dal_t-\dalhad}.
      \ee
      We treat $\alpha_0=1/137.035999074(44)$~\cite{Beringer:1900zz,constants} as a constant, since it is extraordinarily well measured. The contribution from leptons $\dal_\ell$ and the top quark $\dal_t$ are perturbatively calculable and known very accurately, so the uncertainty in $\almz$ essentially comes from the incalculable light hadron contribution $\dalhad$, which is extracted from low energy $e^+ e^- \to$ hadrons data via dispersion relations~\cite{Beringer:1900zz}. For simplicity, we will occasionally (especially in subscripts) drop the scale ``$(m_Z)$'' in $\alsmz$ and $\almz$, and write $\dalhad$ as $\dal$ in the following.

  \item $Z$ boson decay observables: total width $\Gamma_Z$, and partial widths into fermions $\Gamma_f\equiv\Gamma(Z\to f\bar f)$. Also we define and use the invisible and hadronic partial widths\footnote{$\Gamma\had$ is not quite the sum of all $\Gamma_q$, as there are $\Ord{\alpha_s^3}$ corrections that cannot be attributed to any $\Gamma_q$~\cite{Chetyrkin:1996ia}. However, these corrections are small, and are neglected in ZFITTER. We will come back to this in Appendix~\ref{app:qcd}.}:
  \be
  \Gamma\inv\equiv3\Gamma_{\nu},\quad \Gamma\had\equiv\Gamma(Z\to\text{hadrons})\simeq\Gamma_u+\Gamma_d+\Gamma_c+\Gamma_s+\Gamma_b.
  \ee
  The ratios of partial widths are defined and also included in our observables list:
      \be
      R_\ell \equiv \frac{\Gamma\had}{\Gamma_\ell}, \qquad R_q \equiv \frac{\Gamma_q}{\Gamma\had},
      \ee
  where $\ell$ and $q$ denote any one of the lepton and quark species, respectively.
  \item $e^+ e^- \to$ hadrons cross section at the $Z$ pole:
  \be
  \sigma\had = 12\pi \frac{\Gamma_e\Gamma\had}{m_Z^2\Gamma_Z^2}.
  \ee

  \item Forward-backward asymmetries for $e^+ e^- \to f\bar f$ at the $Z$ pole:
      \be
      \AFB{f} = \frac{\sigma_{\text{F}}-\sigma_{\text{B}}}{\sigma_{\text{F}}+\sigma_{\text{B}}} = \frac{3}{4}\Asy{e}\Asy{f}.
      \ee
      The asymmetry parameters $\Asy{f}$ are related to the definition of the effective electroweak mixing angle $\sineff{f}$ by
      \be
      \Asy{f} = \frac{2 (1 - 4 |Q_f| \sineff{f})}{1 + (1 - 4 |Q_f| \sineff{f})^2},
      \ee
      where $Q_f$ is the electric charge of fermion $f$.
\end{itemize}
The experimental results for these observables are listed in Table~\ref{table:obs}. For all the $Z$ pole observables, we use the numbers presented in~\cite{ALEPH:2005ab}, which are combinations of various experimental results at LEP and SLC. Among these observables, lepton universality is assumed only for $\sineff{e}$. For $\sineff{e}$, we also list the PDG combination~\cite{Beringer:1900zz} of D0~\cite{Abazov:2011ws} and CDF~\cite{Han:2011vw} results (the second number). $m_W$ from~\cite{Group:2012gb} is the average of LEP2~\cite{Alcaraz:2006mx} and Tevatron~\cite{Group:2012gb} results. $m_H$ is the PDG average~\cite{Beringer:1900zz} of ATLAS~\cite{Aad:2012tfa} and CMS~\cite{Chatrchyan:2012jja} results.

Table~\ref{table:obs} also contains the reference theory values around which we expand, and their percent relative uncertainties. These theory quantities will be introduced and discussed in detail in Section~\ref{sec:recasting}.

\begin{table}
\begin{center}
  \begin{tabular}{|l|r|r|c|}
    \hline
    $\Obs{i}{}$ & $\Obs{i}{expt}$ & $\Obs{i}{ref}$ & $P[\Obs{i}{ref}]$ \\
    \hline
 $m_Z$ [GeV] & 91.1876(21)~\cite{ALEPH:2005ab}           &    91.1876       & \\
 $G_F$ [GeV$^{-2}$] & 1.1663787(6)e-5~\cite{Beringer:1900zz}    &   1.1663787e-5  & \\
 $\dalhad$ & 0.02772(10)~\cite{Beringer:1900zz}             &   0.02772  & \\
 $m_t$ [GeV] & 173.20(87)~\cite{CDF:2013jga}           &    173.20       & \\
 $\alsmz$ & 0.1185(6)~\cite{Beringer:1900zz}              &   0.1185       & \\
 $m_H$ [GeV] & 125.9(4)~\cite{Beringer:1900zz}           &    125.9       & \\
 \hline
 $\almz$ & 7.81592(86)e-3~\cite{Beringer:1900zz}                  &   7.75611e-3  &
  0.01 \\
 $m_W$ [GeV] & 80.385(15)~\cite{Group:2012gb}  &    80.3614       &
  0.01 \\
 $\Gamma_e$ [MeV] & 83.92(12)~\cite{ALEPH:2005ab}         &    83.9818       &
  0.02 \\
 $\Gamma_{\mu}$ [MeV] & 83.99(18)~\cite{ALEPH:2005ab}     &    83.9812       &
  0.02 \\
 $\Gamma_{\tau}$ [MeV] & 84.08(22)~\cite{ALEPH:2005ab}    &    83.7916       &
  0.02 \\
 $\Gamma_b$ [MeV] & 377.6(1.3)~\cite{ALEPH:2005ab}         &    375.918       &
  0.04 \\
 $\Gamma_c$ [MeV] & 300.5(5.3)~\cite{ALEPH:2005ab}         &    299.969       &
  0.06 \\
 $\Gamma\inv$ [GeV] & 0.4974(25)~\cite{ALEPH:2005ab}       &   0.501627       &
  0.02 \\
 $\Gamma\had$ [GeV] & 1.7458(27)~\cite{ALEPH:2005ab}       &    1.74169       &
  0.04\\
 $\Gamma_Z$ [GeV] & 2.4952(23)~\cite{ALEPH:2005ab}         &    2.49507       &
  0.03 \\
 $\sigma\had$ [nb] & 41.541(37)~\cite{ALEPH:2005ab}        &    41.4784       &
  0.01\\
 $R_e$ & 20.804(50)~\cite{ALEPH:2005ab}                    &    20.7389       &
  0.03 \\
 $R_{\mu}$ & 20.785(33)~\cite{ALEPH:2005ab}                &    20.7391       &
  0.03 \\
 $R_{\tau}$ & 20.764(45)~\cite{ALEPH:2005ab}               &    20.7860       &
  0.03 \\
 $R_b$ & 0.21629(66)~\cite{ALEPH:2005ab}                    &   0.215835       &
  0.02 \\
 $R_c$ & 0.1721(30)~\cite{ALEPH:2005ab}                    &   0.172229       &
  0.01 \\
 $\sineff{e}$ & 0.23153(16)~\cite{ALEPH:2005ab} & 0.231620 &
  0.04 \\
  & 0.23200(76)~\cite{Beringer:1900zz}  &          &  \\
 $\sineff{b}$ & 0.281(16)~\cite{ALEPH:2005ab}             &   0.232958       &
  0.03 \\
 $\sineff{c}$ & 0.2355(59)~\cite{ALEPH:2005ab}             &   0.231514       &
  0.04 \\
 $\Asy{e}$ & 0.1514(19)~\cite{ALEPH:2005ab}                &   0.146249       &
  0.44 \\
 $\Asy{b}$ & 0.923(20)~\cite{ALEPH:2005ab}                &   0.934602       &
  0.00 \\
 $\Asy{c}$ & 0.670(27)~\cite{ALEPH:2005ab}                &   0.667530       &
  0.04 \\
 $\AFB{e}$ & 0.0145(25)~\cite{ALEPH:2005ab}                &   0.0160415  &
  0.88 \\
 $\AFB{b}$ & 0.0992(16)~\cite{ALEPH:2005ab}                &   0.102513       &
  0.44 \\
 $\AFB{c}$ & 0.0707(35)~\cite{ALEPH:2005ab}                &   0.0732191  &
  0.48 \\
    \hline
  \end{tabular}
\end{center}
\caption{The list of observables, their experimental and reference values, and percent relative uncertainties. We set $\Obs{i'}{ref}=\Obs{i'}{expt}$ for the input observables, and calculate $\Obs{i}{ref}$ for the other observables. The percent relative uncertainty $P[\Obs{i}{ref}]$ is the maximum deviation of $\Obs{i}{}$ from $\Obs{i}{ref}$ in units of percentage when the input observables are varied within experimental errors; see \eq{pru} (e.g.\ $m_W$ deviates from $\left[m_W\right]^{\text{ref}}$ by at most $0.01\%$).}\label{table:obs}
\end{table}

\section{The formalism}\label{sec:formalism}

\subsection{Expansion about reference point}
Let us denote the set of SM parameters by $\{\p{k'}{}\}$, and the set of SM observables by $\{\Obs{i}{}\}$. The theoretical prediction for each observable can be calculated in the SM as a function of all parameters:
\be
\Obs{i}{th} = \Obs{i}{SM} (\{\p{k'}{}\}).
\ee
The notation here is that primed roman indices run from 1 to $N_p$, the number of SM parameters, while unprimed ones run from 1 to $N_O$, the number of observables under consideration. Note that $N_p$ is finite, while $N_O$ can presumably be infinite (we must at least have $N_O>N_p$ in order to test any theory). The analysis in this paper is done with $N_p=6$ and $N_O=31$, with $\{\p{k'}{}\}$ given in \eq{SMparam} and $\{\Obs{i}{}\}$ listed in Table~\ref{table:obs}.

Next, suppose we want to study some new physics model beyond the SM, which contains a set of new parameters collectively denoted as $\p{}{NP}$ (``NP'' for ``new physics''). Then at least some $\Obs{i}{th}$ will receive new contribution. We expect such new contribution to be small, in the light of apparently good agreement between SM predictions and precision electroweak data. We can thus write
\beq{OthNP}
\Obs{i}{th} = \Obs{i}{SM} (\{\p{k'}{}\}) + \del{NP}\Obs{i}{}(\{\p{k'}{}\}, \p{}{NP}).
\eeq
We wish to decide whether the new physics model is compatible with precision electroweak data, i.e.\ whether the $\Obs{i}{th}$ predicted by \eq{OthNP} are compatible with the experimentally measured values $\Obs{i}{expt}$.

One common misconception in such analysis is that a new physics model would be ruled out if, for some very precisely measured observables, e.g.\ $G_F^{\text{expt}}=1.1663787(6)\times10^{-5}\text{ GeV}^{-2}$, the new physics contribution $\del{NP}\Obs{i}{}$ exceeds the experimental error. The point is that the SM parameters $\{\p{k'}{}\}$ are not directly measured experimentally. Rather, in testing the SM, we adjust $\{\p{k'}{}\}$ and see that for some choice of all parameters $\{\p{k'}{ref}\}$, all $\Obs{i}{SM}$ agree well with $\Obs{i}{expt}$. In the presence of new physics, we should do the same thing, and will typically arrive at a different choice of $\{\p{k'}{ref}\}$, and hence different $\Obs{i}{SM}$, which may allow the new physics model to survive (in some regions of parameter space spanned by $\p{}{NP}$) despite a large $\del{NP}\Obs{i}{}$.

The statements above are made more precise by the $\chi^2$ analysis, which is the standard way of doing precision electroweak analysis. With correlations among the observables ignored, and experimental errors assumed larger than theoretical errors, the $\chi^2$ function is defined by
\be
\chi^2 (\{\p{k'}{}\}, \p{}{NP}) = \sum_i \left[\frac{\Obs{i}{th}(\{\p{k'}{}\}, \p{}{NP}) - \Obs{i}{expt}}{\Delta\Obs{i}{expt}}\right]^2,
\ee
where $\Delta\Obs{i}{expt}$ are the experimental uncertainties of the observables. To decide whether some $\p{}{NP}$ in the new physics model parameter space survives precision tests, we vary $\{\p{k'}{}\}$ to minimize the $\chi^2$ function to find the best fit to experimental data, and see if this minimum $\chi^2$ is small enough. A good discussion of how to interpret the statistics of the $\chi^2$ distribution can be found in~\cite{Beringer:1900zz}.

In principle, one can calculate $\Obs{i}{th}$ each time a different $\{\p{k'}{}\}$ is chosen in this minimization procedure. But in practice, we can do it once and for all by carrying out an expansion about some reference point in the SM parameter space $\{\p{k'}{ref}\}$. Such an expansion is useful because precision data does not allow much variation in each parameter. Thus, let's choose some $\{\p{k'}{ref}\}$ that lead to good agreement between $\Obs{i}{SM}$ and $\Obs{i}{expt}$, and write
\beq{OSMexpand}
\Obs{i}{SM} (\{\p{k'}{}\}) = \Obs{i}{ref} + \sum_{k'} \frac{\partial\Obs{i}{SM}}{\partial\p{k'}{}} (\p{k'}{} - \p{k'}{ref}) + \dots
\eeq
where $\Obs{i}{ref}\equiv\Obs{i}{SM}(\{\p{k'}{ref}\})$, and the partial derivatives are taken at $\p{k'}{}=\p{k'}{ref}$ (this will be implicitly assumed in the following). Alternatively, define
\be
\delb{SM}\Obs{i}{}(\{\p{k'}{}\}) \equiv \frac{\Obs{i}{SM}(\{\p{k'}{}\}) - \Obs{i}{ref}}{\Obs{i}{ref}} \text{ ,} \qquad
\db\p{k'}{} \equiv \frac{\p{k'}{} - \p{k'}{ref}}{\p{k'}{ref}} \text{ ,} \qquad
G_{ik'} \equiv \frac{\p{k'}{ref}}{\Obs{i}{ref}} \frac{\partial\Obs{i}{SM}}{\partial\p{k'}{}}.
\ee
Then we have a more concise expression for \eq{OSMexpand}:
\beq{G}
\delb{SM}\Obs{i}{} = \sum_{k'} G_{ik'} \db\p{k'}{} + \dots
\eeq
Here $\delb{}$ means ``fractional shift from the reference value'', and the superscript on $\delb{SM}\Obs{i}{}$ indicates the shift comes from shifts in SM parameters. Ignoring higher order terms in the expansion, the constant $G_{ik'}$ is the fractional change in $\Obs{i}{SM}$ caused by the fractional change in $\p{k'}{}$, and hence characterizes the sensitivity of the $i$th SM observable (as calculated in the SM) to the $k'$th SM parameter.

In the presence of perturbative new physics contributions, let's define
\beq{xidef}
\db\Obs{i}{th}(\{\p{k'}{}\},\p{}{NP}) \equiv \frac{\Obs{i}{th}(\{\p{k'}{}\},\p{}{NP}) - \Obs{i}{ref}}{\Obs{i}{ref}} \text{ ,} \qquad
\xi_i(\{\p{k'}{}\}, \p{}{NP}) \equiv \frac{\del{NP}\Obs{i}{}(\{\p{k'}{}\}, \p{}{NP})}{\Obs{i}{ref}}.
\eeq
Then \eq{OthNP} can be expanded as, to first order,
\beq{Gxi}
\db\Obs{i}{th} = \delb{SM}\Obs{i}{} + \xi_i = \sum_{k'} G_{ik'} \db\p{k'}{} + \xi_i.
\eeq
The calculation of $\Obs{i}{th}$ and hence $\chi^2$ is then facilitated if we have at hand the constants $\p{k'}{ref}$, $\Obs{i}{ref}$ and $G_{ik'}$.

\subsection{Recasting observables in terms of observables}\label{sec:recasting}

The approach above is indirect, in the sense that the input of the analysis, the parameters $\{\p{k'}{}\}$, are not directly measurable -- only $\{\Obs{i}{}\}$ are well-defined observables. We can do better if we use $N_p$ very well measured observables $\{\Obs{i'}{}\}$ as input. Note that primed indices, which run from 1 to $N_p$, are used for input observables. Inverting the functions $\Obs{i'}{SM}(\{\p{k'}{}\})$, we can express other observables as functions of these input observables. Then it is immediately clear from $\Obs{i'}{expt}$ and $\Delta\Obs{i'}{expt}$ what reference values for the input we should use, and by how much they are allowed to vary. In our analysis, $N_p=6$, and a convenient choice for the 6 input observables is
\beq{input1}
\{\Obs{i'}{}\}=\{ m_Z,\; G_F,\; \dalhad,\; m_t,\; \alsmz,\; m_H \}.
\eeq
The reference values for these input observables are taken to be the central values experimentally measured; see Table~\ref{table:obs}. All other observables are output observables, and their reference values $\Obs{i}{ref}$ are evaluated at $\Obs{i'}{}=\Obs{i'}{ref}$ with the help of ZFITTER. See Appendix~\ref{app:zfitter} for technical details.

We also show in Table~\ref{table:obs} the ``percent relative uncertainties'' $P[\Obs{i}{ref}]$, defined as the maximum value of
\beq{pru}
100 \left|\frac{\Obs{i}{SM}(\{\Obs{i'}{}\}) - \Obs{i}{ref}}{\Obs{i}{ref}}\right|
\eeq
when all $\{\Obs{i'}{}\}$ are varied in their 1$\sigma$ range around $\{\Obs{i'}{expt}\}$. We do not distinguish between positive and negative relative uncertainties because, as we have checked, the asymmetry in the uncertainties for all observables considered here are very small.

To work out the expansion about the reference point, we assume the input observables $\{\Obs{i'}{}\}$ are the first $N_p$ observables in the list $\{\Obs{i}{}\}$. Then we can simply invert the first $N_p$ equations in \eq{G}. To first order,
\be
\delb{SM}\Obs{i'}{} = \sum_{k'} G_{i'k'} \db\p{k'}{} = \sum_{k'} \widetilde{G}_{i'k'} \db\p{k'}{} \quad \Rightarrow \quad \db\p{k'}{} = \sum_{i'} (\widetilde{G}^{-1})_{k'i'} \delb{SM}\Obs{i'}{}.
\ee
Note that $G$ is a $N_O\times N_p$ matrix, while $\widetilde{G}$ is the upper $N_p\times N_p$ block of $G$. Then \eq{G} suggests
\beq{dbOSM}
\delb{SM}\Obs{i}{} = \sum_{k',i'} G_{ik'} (\widetilde{G}^{-1})_{k'i'} \delb{SM}\Obs{i'}{} \equiv \sum_{i'} c_{ii'}\delb{SM}\Obs{i'}{},
\eeq
where we have defined
\beq{ciip}
c_{ii'} \equiv \sum_{k'} G_{ik'} (\widetilde{G}^{-1})_{k'i'} = \frac{\Obs{i'}{ref}}{\Obs{i}{ref}} \frac{\partial\Obs{i}{SM}}{\partial\Obs{i'}{SM}}.
\eeq
\eq{dbOSM} expresses the shift in any observable in terms of shifts in the input observables, as calculated in the SM. Notably, the upper $N_p\times N_p$ block of the $N_O\times N_p$ matrix $c$ is the identity matrix, i.e.\ $c_{j'i'}=\delta_{j'i'}$. For $i>N_p$, i.e.\ the output observables, the calculation of $c_{ii'}$ is nontrivial. We present in Table~\ref{table:ciip} the results for these expansion coefficients for the observables discussed in Section~\ref{sec:SMpO}, which we calculate using ZFITTER. These coefficients are useful not only because they facilitate the calculation of SM observables. They also give us information on the sensitivity of the calculated observables to each input observable.

\begin{table}
\begin{center}
  \begin{tabular}{|l|rrrrrr|}
    \hline
    $\Obs{i}{}$ & $c_{i,m_Z}$ & $c_{i,G_F}$ & $c_{i,\dal}$ & $c_{i,m_t}$ & $c_{i,\alpha_s}$ & $c_{i,m_H}$ \\
    \hline
    $m_Z$ & 1 & 0 & 0 & 0 & 0 & 0 \\
    $G_F$ & 0 & 1 & 0 & 0 & 0 & 0 \\
    $\dalhad$ & 0 & 0 & 1 & 0 & 0 & 0 \\
    $m_t$ & 0 & 0 & 0 & 1 & 0 & 0 \\
    $\alsmz$ & 0 & 0 & 0 & 0 & 1 & 0 \\
    $m_H$ & 0 & 0 & 0 & 0 & 0 & 1 \\
    \hline
 $\almz$ &              4.796e-3  &    0       &   0.02946  &   1.541e-4  &  -1.007e-5  &    0      \\
 $m_W$ &                 1.427       &   0.2201       &  -6.345e-3  &   0.01322  &  -9.599e-4  &  -7.704e-4 \\
 $\Gamma_e$ &            3.377       &    1.198       &  -5.655e-3  &   0.01883  &  -1.253e-3  &  -7.924e-4 \\
 $\Gamma_{\mu}$ &        3.377       &    1.198       &  -5.655e-3  &   0.01883  &  -1.253e-3  &  -7.924e-4 \\
 $\Gamma_{\tau}$ &       3.383       &    1.198       &  -5.668e-3  &   0.01884  &  -1.254e-3  &  -7.931e-4 \\
 $\Gamma_b$ &            3.844       &    1.411       &  -0.01227  &  -0.01267  &   0.03672  &  -1.057e-3 \\
 $\Gamma_c$ &            4.151       &    1.590       &  -0.01721  &   0.02751  &   0.05046  &  -1.394e-3 \\
 $\Gamma\inv$ &          2.996       &    1.006       &   5.635e-5  &   0.01567  &  -9.967e-4  &  -4.873e-4 \\
 $\Gamma\had$ &          3.938       &    1.476       &  -0.01393  &   0.01578  &   0.03690  &  -1.204e-3 \\
 $\Gamma_Z$ &            3.692       &    1.353       &  -0.01028  &   0.01607  &   0.02543  &  -1.019e-3 \\
 $\sigma\had$ &         -2.069       &  -0.03281  &   9.806e-4  &   2.476e-3  &  -0.01522  &   4.057e-5 \\
 $R_e$ &                0.5608       &   0.2780       &  -8.272e-3  &  -3.045e-3  &   0.03815  &  -4.120e-4 \\
 $R_{\mu}$ &            0.5608       &   0.2780       &  -8.272e-3  &  -3.045e-3  &   0.03815  &  -4.120e-4 \\
 $R_{\tau}$ &           0.5554       &   0.2776       &  -8.259e-3  &  -3.053e-3  &   0.03816  &  -4.113e-4 \\
 $R_b$ &               -0.09434  &  -0.06530  &   1.652e-3  &  -0.02845  &  -1.782e-4  &   1.477e-4 \\
 $R_c$ &                0.2133       &   0.1135       &  -3.284e-3  &   0.01173  &   0.01356  &  -1.898e-4 \\
 $\sineff{e}$ &         -2.818       &   -1.423       &   0.04203  &  -0.02330  &   1.796e-3  &   2.195e-3 \\
 $\sineff{b}$ &         -2.823       &   -1.417       &   0.04204  &  -6.914e-3  &   1.201e-3  &   2.116e-3 \\
 $\sineff{c}$ &         -2.819       &   -1.423       &   0.04202  &  -0.02331  &   1.795e-3  &   2.194e-3 \\
 $\Asy{e}$ &             35.13       &    17.74       &  -0.5239       &   0.2905       &  -0.02239  &  -0.02737 \\
 $\Asy{b}$ &            0.4525       &   0.2271       &  -6.737e-3  &   1.108e-3  &  -1.924e-4  &  -3.390e-4 \\
 $\Asy{c}$ &             3.386       &    1.710       &  -0.05048  &   0.02800  &  -2.156e-3  &  -2.636e-3 \\
 $\AFB{e}$ &             70.27       &    35.48       &   -1.048       &   0.5810       &  -0.04479  &  -0.05473 \\
 $\AFB{b}$ &             35.59       &    17.97       &  -0.5306       &   0.2916       &  -0.02259  &  -0.02771 \\
 $\AFB{c}$ &             38.52       &    19.45       &  -0.5744       &   0.3185       &  -0.02455  &  -0.03000 \\
    \hline
  \end{tabular}
\end{center}
\caption{Expansion coefficients, as defined in \eq{ciip}, calculated in the basis of input observables containing $\dalhad$. These encode the dependence of the output observables on each input observable, and can be used to easily calculate the deviation of the theory prediction of the observables from their reference values via \eq{dbOth}, including new physics contributions.}\label{table:ciip}
\end{table}

In the presence of new physics, \eq{Gxi} becomes
\beq{dbOth}
\db\Obs{i}{th} = \sum_{i'} c_{ii'}\delb{SM}\Obs{i'}{} + \xi_i = \sum_{i'} c_{ii'} (\db\Obs{i'}{th} - \xi_{i'}) + \xi_i = \sum_{i'} c_{ii'}\db\Obs{i'}{th} + \delb{NP}\Obs{i}{},
\eeq
where
\beqa{dbONP}
\delb{NP}\Obs{i}{} &\equiv&\xi_i - \sum_{i'} c_{ii'} \xi_{i'}\nonumber\\
&=& \xi_i - c_{i,m_Z} \xi_{m_Z} - c_{i,G_F} \xi_{G_F} - c_{i,\dal} \xi_{\dal} - c_{i,m_t} \xi_{m_t} - c_{i,\alpha_s} \xi_{\alpha_s} - c_{i,m_H} \xi_{m_H}.
\eeqa
\eq{dbOth} expresses the shift in any observable in terms of shifts in the input observables and new physics effects. Note that for the input observables, since $c_{j'i'}=\delta_{j'i'}$, \eq{dbONP} indicates $\delb{NP}\Obs{i'}{}=0$, and \eq{dbOth} trivially becomes $\db\Obs{i'}{th}=\db\Obs{i'}{th}$. This is forced to be true in our formalism, where $\Obs{i'}{th}$ are inputs of the analysis, independent of new physics. Of course, new physics does contribute $\xi_{i'}$ to the calculation of $\Obs{i'}{th}$, but as we decide to use some particular values for the input $\Obs{i'}{th}$ to be consistent with $\Obs{i'}{expt}$ (which are extraordinarily well measured), we find ourselves adjusting the SM parameters to compensate for $\xi_{i'}$. This adjustment gets propagated into the shift in $\Obs{i}{th}$ due to new physics for $i>N_p$. As a result, \eq{dbONP} shows that for the output observables, $\delb{NP}\Obs{i}{}$ is not simply $\xi_i$, but is related to $\xi_{i'}$ for all input observables.

To close this subsection we remark on the calculation of $\xi_i$. In practice this is done at tree-level or one-loop-level, if we are only interested in constraining a new physics model at percentage level accuracy. Also, the definition of $\xi_i$, \eq{xidef}, instructs us to calculate them in terms of Lagrangian parameters, which can then be eliminated in favor of input observables using the tree-level relations between the two. This does not conflict with the ``precision'' part of the analysis, since we are doing two different perturbative expansions in the calculation: the expansion in SM couplings, and the expansion in new physics effects. Since new physics makes tiny contributions to $\Obs{i}{th}$, to discern them we have to calculate the SM part as precisely as possible, carrying out the expansion in SM couplings to as high order as possible. On the other hand, in most cases the new physics contributions $\xi_i$ need not be calculated beyond leading order, since they are already very small. We will see explicitly how the reasoning above works out in specific examples in Section~\ref{sec:dim6}.

\subsection{Beyond first order}\label{sec:beyond}

The above perturbative expansion carried out to first order is expected to be sufficient for the purpose of precision electroweak analysis, since we have chosen a very well-measured set of input observables, so that the expansion parameters $\db\Obs{i'}{th}$ are tiny. The impact of higher order terms in the expansion can be seen from the sensitivity of the expansion coefficients $c_{ii'}$ to the choice of reference values for the input observables $\Obs{i'}{ref}$. In Table~\ref{table: Pciip} we show the percent relative uncertainties for $c_{ii'}$, defined similarly to \eq{pru}.

Alternatively, without varying $\Obs{i'}{ref}$, we can explicitly write down the next order terms in the expansion:
\beq{higherorder}
\delb{SM}\Obs{i}{} = \sum_{i'} c_{ii'}\delb{SM}\Obs{i'}{} + \frac{1}{2!}\sum_{i'j'} c_{ii'j'}\delb{SM}\Obs{i'}{}\delb{SM}\Obs{j'}{} + \dots \equiv \sum_{i'} (c_{ii'}+\Delta c_{ii'}) \delb{SM}\Obs{i'}{} + \dots
\eeq
where
\beq{ciipjp}
c_{ii'j'} \equiv \frac{\Obs{i'}{ref}\Obs{j'}{ref}}{\Obs{i}{ref}} \frac{\partial^2\Obs{i}{SM}}{\partial\Obs{i'}{SM}\partial\Obs{j'}{SM}}.
\eeq
Then the size of second order terms in \eq{higherorder} compared with the first order term is characterized by the ratio
\beq{riip}
\left|\frac{\Delta c_{ii'}}{c_{ii'}}\right| = \left|\frac{\sum_{j'} c_{ii'j'} \delb{SM}\Obs{j'}{}}{2 c_{ii'}}\right| \le \frac{\sum_{j'} |c_{ii'j'}| |\delb{SM}\Obs{j'}{}|}{2 |c_{ii'}|} \equiv 0.01r_{ii'}.
\eeq
We show in Table~\ref{table:ratio} the $r_{ii'}$ calculated with $\delb{SM}\Obs{j'}{} = \Delta\Obs{j'}{expt}/\Obs{j'}{ref}$. The results follow a similar pattern as in Table~\ref{table: Pciip}.

Tables~\ref{table: Pciip} and~\ref{table:ratio} both show that the uncertainties on the observables calculations are negligible due to uncertainty in the first-order expansion coefficient $c_{ii'}$'s. Most entries manifestly demonstrate this with values of less than $1\%$ corrections to the first-order coefficients that are already governing less than $1\%$ shifts in the observables due to the small uncertainties of the input observables to the calculation (see Table~\ref{table:obs}). Only in a couple of places does the uncertainty reach more than $1\%$, but the final uncertainty on the observables themselves is of course significantly lower than that. To illustrate this, let us consider the largest $P[c_{ii'}]$ in Table~\ref{table: Pciip}, $P[c_{R_b,\alpha_s}]$, which is the uncertainty in the expansion coefficient of $\alpha_s-\alpha_s^{\text{ref}}$ in the computation for $R_b$. It yields an uncertainty on $R_b$ of
\bea
\Delta R_b &\simeq& R_b^{\text{ref}} \left|22\% \times c_{R_b,\alpha_s}\times \db\alpha_s\right|\nonumber\\
 &\simeq& 0.216\, (0.22\times0.0002\times0.005) \simeq 5\times10^{-8},
 \eea
 which is much smaller than the experimental uncertainty of $7\times 10^{-4}$. Therefore, in practice this $22\%$ uncertainty does not concern us, and we can be confident that the first-order expansion expressions are sufficient for any precision electroweak analysis given the current uncertainties in observables.

However, this large uncertainty in $c_{R_b,\alpha_s}$, plus the intuitively unexpected large difference in $c_{\Gamma_q,\alpha_s}$ among different quarks (see Table~\ref{table:correct} in Appendix~\ref{app:qcd}), inspire us to examine closely the calculation of the QCD corrections to $Z$ decay. We will address this issue and explain these features in Appendix~\ref{app:qcd}.

\begin{table}
\begin{center}
  \begin{tabular}{|l|rrrrrr|}
    \hline
    $\Obs{i}{}$ & $P[c_{i,m_Z}]$ & $P[c_{i,G_F}]$ & $P[c_{i,\dal}]$ & $P[c_{i,m_t}]$ & $P[c_{i,\alpha_s}]$ & $P[c_{i,m_H}]$\\
    \hline
 $\almz$ &              0.05  &   - &  0.37 &  1.19 &  1.64 &   - \\
 $m_W$ &                0.02  &  0.05 &  0.44 &  0.87 &  1.20 &  0.23 \\
 $\Gamma_e$ &           0.04  &  0.07 &  0.42 &  1.09 &  1.53 &  0.60 \\
 $\Gamma_{\mu}$ &       0.04  &  0.07 &  0.42 &  1.09 &  1.53 &  0.60 \\
 $\Gamma_{\tau}$ &      0.04  &  0.07 &  0.42 &  1.09 &  1.53 &  0.60 \\
 $\Gamma_b$ &           0.01  &  0.02 &  0.43 &  0.96 &  0.41 &  0.27 \\
 $\Gamma_c$ &           0.01  &  0.01 &  0.39 &  0.88 &  0.64 &  0.33 \\
 $\Gamma\inv$ &         0.00  &  0.01 &  0.63 &  1.04 &  1.51 &  0.74 \\
 $\Gamma\had$ &         0.01  &  0.01 &  0.41 &  1.10 &  0.50 &  0.35 \\
 $\Gamma_Z$ &           0.00  &  0.01 &  0.39 &  1.07 &  0.52 &  0.39 \\
 $\sigma\had$ &         0.06  &  2.08 &  2.41 &  1.31 &  0.50 &  2.81 \\
 $R_e$ &                0.31  &  0.32 &  0.69 &  1.40 &  0.47 &  0.36 \\
 $R_{\mu}$ &            0.31  &  0.32 &  0.69 &  1.40 &  0.47 &  0.36 \\
 $R_{\tau}$ &           0.32  &  0.33 &  0.69 &  1.40 &  0.47 &  0.36 \\
 $R_b$ &                0.13 &  0.28 &  0.41 &  0.92 & 22.06 &  0.88 \\
 $R_c$ &                0.12 &  0.14 &  0.41 &  0.87 &  1.26 &  0.35 \\
 $\sineff{e}$ &         0.02 &  0.01 &  0.39 &  0.97 &  1.26 &  0.12 \\
 $\sineff{b}$ &         0.02 &  0.02 &  0.39 &  0.75 &  1.16 &  0.05 \\
 $\sineff{c}$ &         0.02 &  0.01 &  0.39 &  0.97 &  1.26 &  0.12 \\
 $\Asy{e}$ &            0.51 &  0.50 &  0.88 &  1.10 &  1.42 &  0.46 \\
 $\Asy{b}$ &            0.09 &  0.09&  0.46 &  0.80 &  1.21 &  0.11 \\
 $\Asy{c}$ &            0.14 &  0.14 &  0.52 &  1.00 &  1.30 &  0.16 \\
 $\AFB{e}$ &            0.51 &  0.50 &  0.88 &  1.10 &  1.42 &  0.46 \\
 $\AFB{b}$ &            0.50 &  0.49 &  0.88 &  1.10 &  1.42 &  0.46 \\
 $\AFB{c}$ &            0.48 &  0.47 &  0.85 &  1.09 &  1.41 &  0.43 \\
    \hline
  \end{tabular}
\end{center}
\caption{Percent relative uncertainties for the expansion coefficients $c_{ii'}$, with all input observables varied in their 1$\sigma$ range.}\label{table: Pciip}
\end{table}

\begin{table}
\begin{center}
  \begin{tabular}{|l|rrrrrr|}
    \hline
    $\Obs{i}{}$ & $r_{i,m_Z}$ & $r_{i,G_F}$ & $r_{i,\dal}$ & $r_{i,m_t}$ & $r_{i,\alpha_s}$ & $r_{i,m_H}$ \\
    \hline
 $\almz$ &              0.03 &   - &  0.01 &  0.85 &  0.66 &   - \\
 $m_W$ &                0.01 &  0.03 &  0.03 &  0.18 &  0.35 &  0.18 \\
 $\Gamma_e$ &           0.03 &  0.04 &  0.20 &  0.30 &  0.52 &  0.18 \\
 $\Gamma_{\mu}$ &       0.03 &  0.04 &  0.20 &  0.30 &  0.52 &  0.18 \\
 $\Gamma_{\tau}$ &      0.03 &  0.04 &  0.20 &  0.30 &  0.52 &  0.18 \\
 $\Gamma_b$ &           0.02 &  0.02 &  0.04 &  0.24 &  0.10 &  0.07 \\
 $\Gamma_c$ &           0.02 &  0.03 &  0.02 &  0.21 &  0.09 &  0.16 \\
 $\Gamma\inv$ &         0.01 &  0.01 &  0.12 &  0.27 &  0.51 &  0.21 \\
 $\Gamma\had$ &         0.02 &  0.02 &  0.02 &  0.29 &  0.04 &  0.14 \\
 $\Gamma_Z$ &           0.02 &  0.02 &  0.02 &  0.29 &  0.05 &  0.13 \\
 $\sigma\had$ &         0.03 &  1.04 &  1.02 &  0.39 &  0.02 &  1.49 \\
 $R_e$ &                0.17 &  0.17 &  0.17 &  0.46 &  0.02 &  0.31 \\
 $R_{\mu}$ &            0.17 &  0.17 &  0.17 &  0.46 &  0.02 &  0.31 \\
 $R_{\tau}$ &           0.17 &  0.17 &  0.17 &  0.46 &  0.02 &  0.31 \\
 $R_b$ &                0.05 &  0.13 &  0.05 &  0.20 & 10.69 &  0.59 \\
 $R_c$ &                0.06 &  0.07 &  0.05 &  0.19 &  0.38 &  0.31 \\
 $\sineff{e}$ &         0.03 &  0.02 &  0.03 &  0.24 &  0.38 &  0.19 \\
 $\sineff{b}$ &         0.03 &  0.02 &  0.03 &  0.13 &  0.34 &  0.17 \\
 $\sineff{c}$ &         0.03 &  0.02 &  0.03 &  0.24 &  0.38 &  0.19 \\
 $\Asy{e}$ &            0.04 &  0.03 &  0.04 &  0.24 &  0.38 &  0.20 \\
 $\Asy{b}$ &            0.04 &  0.04 &  0.05 &  0.14 &  0.35 &  0.18 \\
 $\Asy{c}$ &            0.05 &  0.05 &  0.06 &  0.24 &  0.39 &  0.20 \\
 $\AFB{e}$ &            0.18 &  0.19 &  0.18 &  0.42 &  0.55 &  0.37 \\
 $\AFB{b}$ &            0.03 &  0.03 &  0.04 &  0.24 &  0.38 &  0.19 \\
 $\AFB{c}$ &            0.00 &  0.01 &  0.01 &  0.23 &  0.37 &  0.19 \\
    \hline
  \end{tabular}
\end{center}
\caption{The $r_{ii'}$'s defined in \eq{riip}, characterizing the ratios of second-order vs.\ first-order terms in the expansion (in units of percentage).}\label{table:ratio}
\end{table}

\subsection{Change of basis}\label{sec:changebasis}
Our choice of input observables as in \eq{input1} is convenient for the calculation of expansion coefficients in ZFITTER. In principle, any set of $N_p=6$ independent observables can serve as input, though we should better choose those most precisely measured observables to minimize the uncertainty due to higher order terms in the expansion. In this respect, an equally good choice as \eq{input1} could be
\beq{input2}
\{\Obs{i'}{}\}=\{ m_Z,\; G_F,\; \almz,\; m_t,\; \alsmz,\; m_H \},
\eeq
since essentially all the uncertainty in $\almz$ comes from $\dalhad$. This basis may be preferable in practice, since it is often more convenient to do calculations with $\almz$, rather than $\dalhad$, as input. In this subsection we derive the rules for translating the expansion coefficients $c_{ii'}$, which are calculated in the basis \eq{input1}, into those for the basis \eq{input2}. To avoid confusion, denote the latter by $d_{ii'}$. Also, superscripts ``SM'' will be dropped for simplicity in this subsection.

First, consider $d_{i,\alpha}$. We need to determine the shift in $\Obs{i}{}$ caused by $\db\almz$, with the other 5 input observables held fixed. If we work in the basis \eq{input1}, this shift in $\almz$ is an outcome of the following shift in $\dalhad$ (with other input observables fixed):
\be
\db\dalhad = \left[c_{\alpha,\dal}\right]^{-1} \db\almz.
\ee
And the shift in $\Obs{i}{}$ is
\be
\db\Obs{i}{} = c_{i,\dal} \db\dalhad = c_{i,\dal}\left[c_{\alpha,\dal}\right]^{-1} \db\almz.
\ee
Thus,
\beq{cdconvert1}
d_{i,\alpha} = \frac{\db\Obs{i}{}}{\db\almz} = c_{i,\dal}\left[c_{\alpha,\dal}\right]^{-1}.
\eeq

Next, consider $d_{ii'}$ for $i'\ne\almz$. Take $d_{i,m_Z}$ as an example. We need to shift $m_Z$ while keeping other observables in \eq{input2}, including $\almz$, fixed, and find the resulting shift in $\Obs{i}{}$. Working in the basis \eq{input1}, we can do this in two steps. First, shift $m_Z$ by $\db m_Z$. As a result,
\be
\db\Obs{i}{} = c_{i,m_Z} \db m_Z \text{ ,}\qquad \db\almz = c_{\alpha,m_Z} \db m_Z.
\ee
Second, shift $\dalhad$ by
\be
\db\dalhad = -\left[c_{\alpha,\dal}\right]^{-1} c_{\alpha,m_Z} \db m_Z.
\ee
As a result,
\bea
\db\Obs{i}{} &=& c_{i,\dal} \db\dalhad = -c_{i,\dal}\left[c_{\alpha,\dal}\right]^{-1} c_{\alpha,m_Z} \db m_Z,\\
\db\almz &=& c_{\alpha,\dal} \db\dalhad = -c_{\alpha,m_Z} \db m_Z.
\eea
The effect of both steps is to hold all observables in \eq{input2} other than $m_Z$, in particular $\almz$, fixed. And we get the desired result
\beq{cdconvert2}
d_{i,m_Z} = \frac{\db\Obs{i}{}}{\db m_Z} = c_{i,m_Z} - c_{i,\dal}\left[c_{\alpha,\dal}\right]^{-1} c_{\alpha,m_Z}.
\eeq

As a special case, \eqs{cdconvert1}{cdconvert2} also hold for $i=\dalhad$:
\bea
d_{\dal,\alpha} &=& \left[c_{\alpha,\dal}\right]^{-1},\\
d_{\dal,m_Z} &=& -\left[c_{\alpha,\dal}\right]^{-1} c_{\alpha,m_Z},
\eea
where we have used $c_{\dal,\dal}=1$, $c_{\dal,m_Z}=0$.

In the basis \eq{input2}, the theory predictions for the observables (with respect to the reference values) are calculated from
\beq{dbOthd}
\db\Obs{i}{th} = \sum_{i'} d_{ii'}\db\Obs{i'}{th} + \delb{NP}\Obs{i}{},
\eeq
where
\bea
\delb{NP}\Obs{i}{} &\equiv&\xi_i - \sum_{i'} d_{ii'} \xi_{i'}\nonumber\\
&=& \xi_i - d_{i,m_Z} \xi_{m_Z} - d_{i,G_F} \xi_{G_F} - d_{i,\alpha} \xi_{\alpha} - d_{i,m_t} \xi_{m_t} - d_{i,\alpha_s} \xi_{\alpha_s} - d_{i,m_H} \xi_{m_H}.
\eea
We list the expansion coefficients $d_{ii'}$, as calculated from \eqs{cdconvert1}{cdconvert2}, in Table~\ref{table:diip}.

\begin{table}
\begin{center}
  \begin{tabular}{|l|rrrrrr|}
    \hline
    $\Obs{i}{}$ & $d_{i,m_Z}$ & $d_{i,G_F}$ & $d_{i,\alpha}$ & $d_{i,m_t}$ & $d_{i,\alpha_s}$ & $d_{i,m_H}$ \\
    \hline
    $m_Z$ & 1 & 0 & 0 & 0 & 0 & 0 \\
    $G_F$ & 0 & 1 & 0 & 0 & 0 & 0 \\
    $\almz$ & 0 & 0 & 1 & 0 & 0 & 0 \\
    $m_t$ & 0 & 0 & 0 & 1 & 0 & 0 \\
    $\alsmz$ & 0 & 0 & 0 & 0 & 1 & 0 \\
    $m_H$ & 0 & 0 & 0 & 0 & 0 & 1 \\
    \hline
 $\dalhad$ &      -0.1628       &    0       &    33.94       &   -5.232e-3  &  3.417e-4  &    0      \\
 $m_W$ &                 1.428       &   0.2201       &  -0.2154       &   0.01325  &  -9.621e-4  &  -7.704e-4 \\
 $\Gamma_e$ &            3.378       &    1.198       &  -0.1920       &   0.01886  &  -1.255e-3  &  -7.924e-4 \\
 $\Gamma_{\mu}$ &        3.378       &    1.198       &  -0.1920       &   0.01886  &  -1.255e-3  &  -7.924e-4 \\
 $\Gamma_{\tau}$ &       3.384       &    1.198       &  -0.1924       &   0.01887  &  -1.256e-3  &  -7.931e-4 \\
 $\Gamma_b$ &            3.846       &    1.411       &  -0.4166       &  -0.01260  &   0.03672  &  -1.057e-3 \\
 $\Gamma_c$ &            4.154       &    1.590       &  -0.5842       &   0.02760  &   0.05045  &  -1.394e-3 \\
 $\Gamma\inv$ &          2.996       &    1.006       &   1.913e-3  &   0.01567  &  -9.967e-4  &  -4.873e-4 \\
 $\Gamma\had$ &          3.940       &    1.476       &  -0.4727       &   0.01586  &   0.03690  &  -1.204e-3 \\
 $\Gamma_Z$ &            3.694       &    1.353       &  -0.3490       &   0.01612  &   0.02543  &  -1.019e-3 \\
 $\sigma\had$ &         -2.070       &  -0.03281  &   0.03328  &   2.471e-3  &  -0.01522  &   4.057e-5 \\
 $R_e$ &                0.5622       &   0.2780       &  -0.2807       &  -3.002e-3  &   0.03815  &  -4.120e-4 \\
 $R_{\mu}$ &            0.5622       &   0.2780       &  -0.2807       &  -3.002e-3  &   0.03815  &  -4.120e-4 \\
 $R_{\tau}$ &           0.5568       &   0.2776       &  -0.2803       &  -3.009e-3  &   0.03815  &  -4.113e-4 \\
 $R_b$ &               -0.09461  &  -0.06530  &   0.05608  &  -0.02846  &  -1.777e-4  &   1.477e-4 \\
 $R_c$ &                0.2138       &   0.1135       &  -0.1115       &   0.01174  &   0.01356  &  -1.898e-4 \\
 $\sineff{e}$ &         -2.825       &   -1.423       &    1.426       &  -0.02352  &   1.811e-3  &   2.195e-3 \\
 $\sineff{b}$ &         -2.830       &   -1.417       &    1.427       &  -7.134e-3  &   1.215e-3  &   2.116e-3 \\
 $\sineff{c}$ &         -2.826       &   -1.423       &    1.426       &  -0.02353  &   1.809e-3  &   2.194e-3 \\
 $\Asy{e}$ &             35.22       &    17.74       &   -17.78       &   0.2932       &  -0.02257  &  -0.02737 \\
 $\Asy{b}$ &            0.4536       &   0.2271       &  -0.2287       &   1.143e-3  &  -1.947e-4  &  -3.390e-4 \\
 $\Asy{c}$ &             3.395       &    1.710       &   -1.713       &   0.02827  &  -2.174e-3  &  -2.636e-3 \\
 $\AFB{e}$ &             70.44       &    35.48       &   -35.56       &   0.5865       &  -0.04515  &  -0.05473 \\
 $\AFB{b}$ &             35.67       &    17.97       &   -18.01       &   0.2944       &  -0.02277  &  -0.02771 \\
 $\AFB{c}$ &             38.61       &    19.45       &   -19.50       &   0.3215       &  -0.02475  &  -0.03000 \\
    \hline
  \end{tabular}
\end{center}
\caption{Expansion coefficients calculated in the basis of input observables containing $\almz$, which are derived from the numbers in Table~\ref{table:ciip} by a change of basis described in Section~\ref{sec:changebasis}. These encode the dependence of the output observables on each input observable, and can be used to easily calculate the deviation of the theory prediction of the observables from their reference values via \eq{dbOthd}, including new physics contributions.}\label{table:diip}
\end{table}

\section{New physics examples}\label{sec:NPex}

In this section we present some examples of calculating new physics contributions to electroweak observables, using the formalism developed in Section~\ref{sec:formalism}. We work in the basis \eq{input2}, with $\almz$ as an input observable.

\subsection{Dimension six effective operators}\label{sec:dim6}

The SM, when viewed as an effective field theory below some cutoff scale $\Lambda$, can be supplemented by higher dimensional operators suppressed by powers of $\Lambda$~\cite{Buchmuller:1985jz,Grzadkowski:2010es}, which presumably come from new physics at or above $\Lambda$. Two examples at dimension 6 are:
\beq{OLOH}
\Op_L = \frac{1}{2\Lambda_L^2} \bigl(\bar L \gamma_{\mu} \sigma^a L\bigr)^2, \quad \Op_H = \frac{1}{\Lambda_H^2} \bigl|H^{\dagger} D_{\mu} H\bigr|^2,
\eeq
where $L$ and $H$ are the lepton and Higgs $SU(2)_L$ doublets, respectively, and $\sigma^a$ ($a=1,2,3$) are the Pauli matrices. In this subsection we consider these two operators separately, and illustrate how to use the formalism developed in this paper to work out the precision electroweak constraints on $\Lambda_L$, $\Lambda_H$.

First consider $\Op_L$. At tree level the only nonzero $\xi_i$ at $\Ord{\frac{1}{\Lambda_L^2}}$ is
\be
\xi_{G_F} = \frac{v^2}{\Lambda_L^2} = \frac{1}{\sqrt 2 G_F \Lambda_L^2}.\quad\text{(tree-level)}
\ee
This computation should not be compared with the experimental uncertainty in $G_F$ measurement to get limits on $\Lambda_L^2$.  Rather, we should calculate
\beq{OLdelbNP}
\delb{NP}\Obs{i}{} = \xi_i - d_{i,G_F} \xi_{G_F} \simeq \xi_i - d_{i,G_F}\left(\frac{246\text{ GeV}}{\Lambda_L}\right)^2
\eeq
for all observables using the $d_{i,G_F}$ listed in Table~\ref{table:diip}, and perform a $\chi^2$ analysis. Indeed, \eq{OLdelbNP} gives $\delb{NP} G_F=0$, which is an essential check to the formalism since $G_F$ is an input observable that is by definition set to whatever value we wish it to have. In other words, if new physics does appear to want to shift $G_F$, the parameters in the theory adjust themselves such that the total shift is zero. That is the nature of being a fixed input observable to precision electroweak computations.

Because of the rearrangement of SM parameters due to accommodating the contribution to $G_F$ from new physics, every output observable will feel a shift. For example,
\bea
\delb{NP} m_W &\simeq& - d_{m_W,G_F}\left(\frac{246\text{ GeV}}{\Lambda_L}\right)^2 \simeq -0.220\left(\frac{246\text{ GeV}}{\Lambda_L}\right)^2,\\
\delb{NP}\Asy{e} &\simeq& - d_{\Asy{e},G_F}\left(\frac{246\text{ GeV}}{\Lambda_L}\right)^2 \simeq -17.7\left(\frac{246\text{ GeV}}{\Lambda_L}\right)^2.
\eea
Similar expressions exist for all SM precision electroweak observables. To find limits on $\Lambda_L$ a global $\chi^2$ analysis must be performed, or at least a semi-global $\chi^2$ analysis using the most sensitive observables, such as $\Gamma_e$, $m_W$ and $\sineff{e}$~\cite{Peskin:2001rw}.

Next consider $\Op_H$. In the unitary gauge,
\beq{OH}
H = \frac{1}{\sqrt 2} \left(
\begin{matrix}
0\\
v+h
\end{matrix}
\right) \Rightarrow
\Op_H = \frac{v^2}{2\Lambda_H^2}\left[\frac{1}{2}(\partial_{\mu} h)^2\Bigl(1+\frac{h}{v}\Bigr)^2 + \frac{1}{4}(g_2^2+g_1^2)v^2 Z_{\mu} Z^{\mu} \Bigl(1+\frac{h}{v}\Bigr)^4\right].
\eeq
Noting that $m_Z=\frac{1}{2}\sqrt{g_2^2+g_1^2}v$ at tree level, we have
\be
\xi_{m_Z} = -1+ \Bigl(1+\frac{v^2}{2\Lambda_H^2}\Bigr)^{1/2} \simeq \frac{v^2}{4\Lambda_H^2},\quad\quad
\xi_{m_H} = -1+\Bigl(1+\frac{v^2}{2\Lambda_H^2}\Bigr)^{-1/2} \simeq -\frac{v^2}{4\Lambda_H^2}.\quad\text{(tree-level)}
\ee
The shift in $m_H$ comes from rescaling the field $h$ such that its kinetic term is canonically normalized, as necessitated by the first term in \eq{OH}. To derive constraints on $\Lambda_H$, a $\chi^2$ analysis has to be done, which can be facilitated by the expansion
\be
\delb{NP}\Obs{i}{} = \xi_i - d_{i,m_Z} \xi_{m_Z} - d_{i,m_H} \xi_{m_H} \simeq \xi_i - (d_{i,m_Z} - d_{i,m_H})\left(\frac{123\text{ GeV}}{\Lambda_H}\right)^2.
\ee
Among the output observables in Table~\ref{table:obs}, only those related to $Z$ boson decay have nonzero $\xi_i$ at tree-level due to the shift in $m_Z$:
\bea
\xi_{\Gamma_f} = \xi_{\Gamma\inv} = \xi_{\Gamma\had} = \xi_{\Gamma_Z} = \xi_{m_Z} = \frac{v^2}{4\Lambda_H^2},\\
\xi_{\sigma\had} = -2\xi_{m_Z} = -\frac{v^2}{2\Lambda_H^2}.
\eea
Thus, for example,
\bea
\delb{NP}\Gamma_Z &\simeq& (1- d_{\Gamma_Z,m_Z} + d_{\Gamma_Z,m_H})\left(\frac{123\text{ GeV}}{\Lambda_H}\right)^2 \simeq -2.70 \left(\frac{123\text{ GeV}}{\Lambda_H}\right)^2,\\
\delb{NP} R_b &\simeq& -(d_{R_b,m_Z} - d_{R_b,m_H})\left(\frac{123\text{ GeV}}{\Lambda_H}\right)^2 \simeq 0.0948 \left(\frac{123\text{ GeV}}{\Lambda_H}\right)^2.
\eea

For both operators considered above, the new physics contribution is on the order $\frac{v^2}{\Lambda^2}$. If we were to calculate $\delb{NP}\Obs{i}{}$ to higher order, we would have
\be
\delb{NP}\Obs{i}{}\sim \Ord{\frac{v^2}{\Lambda^2}} \left[1 + \Ord{\frac{v^2}{\Lambda^2}}\right] \left[1+ \Ord{\frac{\alpha_s}{4\pi}}\right].
\ee
Neglecting these higher order corrections will result in errors in the derived constraints on $\Lambda$, typically at the percentage level. However, much effort has been devoted to calculating observables within the SM to a much higher accuracy, and such accuracy is reflected in $\Obs{i}{ref}$ and $d_{ii'}$ presented in this paper. There is no contradiction here, because [recall \eq{dbOthd}]
\be
\Obs{i}{th} = \Obs{i}{ref} (1+\db\Obs{i}{th}) = \Obs{i}{ref} (1+\sum_{i'} d_{ii'}\db\Obs{i'}{th} + \delb{NP}\Obs{i}{}).
\ee
To discern new physics contributions of order $\frac{v^2}{\Lambda^2}$, we must calculate $\Obs{i}{ref}$ and $d_{ii'}$ to a better accuracy, hence the need for higher loop order calculations. The higher order calculation of $\xi_i$, on the other hand, usually does not contribute as much to $\Obs{i}{th}$, because $\delb{NP}\Obs{i}{}$ is $\Ord{\frac{v^2}{\Lambda^2}}$ anyway. In a word, if we only calculate $\xi_i$ (and hence $\delb{NP}\Obs{i}{}$) at tree level, we will constrain new physics models with a few percent uncertainty; but if we didn't calculate $\Obs{i}{ref}$ and $d_{ii'}$ to multi-loop level, we would not be able to constrain them at all!

\subsection{Shifts in $Zb\bar b$ couplings}

Suppose some new physics model shifts the $Z$ boson couplings to left- and right-handed $b$ quarks~\cite{Bamert:1996px}
\be
c_L^b\to c_L^b(1+\varepsilon_L),\quad c_R^b\to c_R^b(1+\varepsilon_R).
\ee
None of the input observables is affected at tree level. Thus, the impact of the shifts of these couplings can be calculated straightforwardly from observables that directly depend on $c_L^b$ and $c_R^b$. The set of observables directly affected include $\Gamma_b$, $\Gamma\had$, $R_{e,\mu,\tau}$, $R_{c,b}$, $\Gamma_Z$, $\sigma\had$, $\Asy{b}$, $\AFB{b}$, and $\sineff{b}$. Their shifts from this new physics contribution can be expressed as
\be
\delb{NP}\Obs{i}{} = \xi_i.
\ee

Let's begin by computing the shift in $\Gamma_b$. At tree level,
$\Gamma_b \propto [(c_L^b)^2+(c_R^b)^2]$, which when expanded leads to the shift $\delb{NP}\Gamma_b = \xi_{\Gamma_b}$, where
\beq{xiGammab}
\xi_{\Gamma_b} = \frac{2(c_L^b)^2}{(c_L^b)^2+(c_R^b)^2}\varepsilon_L + \frac{2(c_R^b)^2}{(c_L^b)^2+(c_R^b)^2}\varepsilon_R \simeq 1.94\,\varepsilon_L + 0.0645\,\varepsilon_R.
\eeq
Knowing this shift in $\Gamma_b$ enables us to simply compute the shift  of other observables that depend on $\Gamma_b$ in terms of $\xi_{\Gamma_b}$:
\bea
\delb{NP}\Gamma\had &=& \delb{NP}R_e = \delb{NP}R_{\mu} = \delb{NP}R_{\tau} = -\delb{NP}R_c = R_b\xi_{\Gamma_b} \simeq 0.216\,\xi_{\Gamma_b},\\
\delb{NP}R_b &=& \delb{NP}\Gamma_b - \delb{NP}\Gamma\had = (1 - R_b)\xi_{\Gamma_b} \simeq 0.784\,\xi_{\Gamma_b},\\
\delb{NP}\Gamma_Z &=& B_b\xi_{\Gamma_b} \simeq 0.151\,\xi_{\Gamma_b},\\
\delb{NP}\sigma\had &=& \delb{NP}\Gamma\had - 2\delb{NP}\Gamma_Z = (R_b - 2 B_b)\xi_{\Gamma_b} \simeq -0.0855\,\xi_{\Gamma_b},
\eea
where $B_b=\Gamma_b/\Gamma_Z$ is the branching ratio of $Z\to b\bar b$.

The asymmetry observables are also affected due to the shift in $\Asy{b}$. At tree level,
\beq{asymb}
\Asy{b} = \frac{(c_L^b)^2-(c_R^b)^2}{(c_L^b)^2+(c_R^b)^2},
\eeq
which leads to a shift $\delb{NP}\Asy{b}=\xi_{\Asy{b}}$, where
\beq{xiAsyb}
\xi_{\Asy{b}} = \frac{4(c_L^b)^2(c_R^b)^2}{(c_L^b)^4-(c_R^b)^4}(\varepsilon_L - \varepsilon_R) \simeq 0.134\,(\varepsilon_L -\varepsilon_R).
\eeq
We can then straightforwardly compute $\delb{NP}\AFB{b}$ and $\delb{NP}\sineff{b}$ in terms of $\xi_{\Asy{b}}$:
\be
\delb{NP}\AFB{b} = \xi_{\Asy{b}},
\ee
and
\be
\delb{NP}\sineff{b}  = \left[\frac{\sineff{b}}{\Asy{b}}\frac{\partial\Asy{b}}{\partial\sineff{b}}\right]^{-1} \xi_{\Asy{b}} = \frac{(1-\frac{4}{3}\sineff{b})[1+(1-\frac{4}{3}\sineff{b})^2]} {-\frac{4}{3}\sineff{b}[1-(1-\frac{4}{3}\sineff{b})^2]} \xi_{\Asy{b}} \simeq  -6.24\, \xi_{\Asy{b}}.
\ee

Thus, $\delb{NP}\Obs{i}{}$ for all observables are expressed in terms of $\xi_{\Gamma_b}$ or $\xi_{\Asy{b}}$, which are simply related to $\varepsilon_L$, $\varepsilon_R$ via \eqs{xiGammab}{xiAsyb}.

\subsection{Shifts in vector boson self-energies}

In many new physics scenarios, there exist exotic states that do not couple directly to SM fermions but have charges under the SM gauge groups. These states affect electroweak observables via shifts in vector boson self-energies~\cite{Peskin:1991sw}. At one-loop level, the dependence of various observables on vector boson self-energies is as follows~\cite{Wells:2005vk}:
\bea
m_Z^2 &=& \tree{m_Z^2}(1+\pizz)\label{mzpi},\\
m_W^2 &=& \tree{m_W^2}(1+\piww),\\
G_F &=& \tree{G_F}(1-\piwwo),\\
\almz &=& \tree{\almz}(1+\piggp),\\
\sineff{f} &=& s^2\Bigl(1-\frac{c}{s}\pigz\Bigr)\label{seffpi},\\
\Gamma_f &=& \tree{\Gamma_f}(1+\pizzp+\frac{1}{2}\pizz+a_f\pigz),\label{Gammafpi}
\eea
where superscripts ``(0)'' denote tree-level values, and $s=\frac{g_1}{\sqrt{g_1^2+g_2^2}}$, $c=\frac{g_2}{\sqrt{g_1^2+g_2^2}}$. We have also defined
\bea
\pizz &\equiv& \frac{\Pi_{ZZ}(m_Z^2)}{m_Z^2},\\
\pizzp &\equiv& \lim_{q^2\to m_Z^2}\frac{\Pi_{ZZ}(q^2) - \Pi_{ZZ}(m_Z^2)}{q^2-m_Z^2},\\
\pigz &\equiv& \frac{\Pi_{\gamma Z}(m_Z^2)}{m_Z^2},\\
\piggp &\equiv& \lim_{q^2\to 0}\frac{\Pi_{\gamma\gamma}(q^2) - \Pi_{\gamma\gamma}(0)}{q^2},\\
\piww &\equiv& \frac{\Pi_{WW}(m_W^2)}{m_W^2},\\
\piwwo &\equiv& \frac{\Pi_{WW}(0)}{m_W^2}.
\eea
The $a_f$ in \eq{Gammafpi} can be derived from
\be
\Gamma_f = \tree{\Gamma_f}(1+\pizzp+\pizz)\frac{1+(1-4|Q_f|\sineff{f})^2}{1+(1-4|Q_f|s^2)^2}
\ee
and \eq{seffpi}. The result is
\beq{afana}
a_f = \frac{8sc|Q_f|(1-4|Q_f|s^2)}{1+(1-4|Q_f|s^2)^2} = 4sc|Q_f|\tree{\Asy{f}}.
\eeq
With $s^2\simeq\sineff{e}=0.231620$, which is good at tree level, we have
\beq{afnum}
a_{\nu}=0,\,\, a_\ell=0.2468,\,\,a_u=0.7505,\,\,a_d=0.5262.
\eeq

With Eqs.~(\ref{mzpi}-\ref{Gammafpi}), it is straightforward to calculate contributions from new physics. Denote the shifts in vector boson self-energies by $\del{NP}\pizz$, etc.; i.e.
\be
\pizz\to\pizz + \del{NP}\pizz,\text{ etc.}
\ee
Note the absence of ``bar'' on $\delta$, since this is the absolute shift, not the fractional shift. Then for the input observables,
\be
\xi_{m_Z} = \frac{1}{2}\del{NP}\pizz,\,\, \xi_{G_F} = -\del{NP}\piwwo,\,\, \xi_{\alpha} = \del{NP}\piggp,\,\, \xi_{m_t} = \xi_{\alpha_s} = \xi_{m_H} = 0.
\ee
These shifts propagate into shifts in the output observables, while leaving the input observables unchanged due to new physics (i.e.\ $\delb{NP}\Obs{i'}{}=0$). The new physics contribution to the output observables can be conveniently expressed as:
\beqa{bdef}
\delb{NP}\Obs{i}{} &=&\xi_i - \sum_{i'} d_{ii'} \xi_{i'}\nonumber\\
&\equiv& b_{i,zz}\del{NP}\pizz + b_{i,zz}^{\prime}\del{NP}\pizzp + b_{i,\gamma z}\del{NP}\pigz + b_{i,\gamma\gamma}^{\prime}\del{NP}\piggp + b_{i,ww}\del{NP}\piww + b_{i,ww}^0\del{NP}\piwwo.
\eeqa
In the following we discuss the calculation of these $b$ coefficients.
\begin{table}
\begin{center}
  \begin{tabular}{|l|rrrrrr|}
    \hline
    $\Obs{i}{}$ & $b_{i,zz}$ & $b_{i,zz}^{\prime}$ & $b_{i,\gamma z}$ & $b_{i,\gamma\gamma}^{\prime}$ & $b_{i,ww}$ & $b_{i,ww}^0$ \\
    \hline
 $m_W$ &                     -0.7140 &  0 &             0 &      0.2154 & 0.5 &  0.2201 \\
 $\Gamma_e$ &            -1.189 &  1 &  0.2468 &      0.1920 &   0  &  1.198 \\
 $\Gamma_{\mu}$ &  -1.189 &  1 &  0.2468 &      0.1920 &   0  &  1.198 \\
 $\Gamma_{\tau}$ &  -1.192 &  1 &  0.2468 &      0.1924 &   0  &  1.198 \\
 $\Gamma_b$ &           -1.423 &  1 &  0.5262 &       0.4166 &   0  &  1.411 \\
 $\Gamma_c$ &            -1.577 &  1 &  0.7505 &       0.5842 &   0  &  1.590 \\
 $\Gamma\inv$ &     -0.9982 &  1 &            0 &  -1.913e-3 &   0  &  1.006 \\
 $\Gamma\had$ &      -1.470 &  1 &  0.6027 &       0.4727 &   0  &  1.476 \\
 $\Gamma_Z$ &           -1.347 &  1 &  0.4420 &       0.3490 &   0  &  1.353 \\
 $\sigma\had$ &       0.03475 &  0 &  -0.03460 &   -0.03328 &   0  &  -0.03281 \\
 $R_e$ &                          -0.2811 &  0 &  0.3559 &       0.2807 &   0  &  0.2780 \\
 $R_{\mu}$ &                -0.2811 &  0 &  0.3559 &       0.2807 &   0  &  0.2780 \\
 $R_{\tau}$ &                -0.2784 &  0 &  0.3559 &       0.2803 &   0  &  0.2776 \\
 $R_b$ &                         0.04731 &  0 &  -0.07647 &   -0.05608 &   0  &  -0.06530 \\
 $R_c$ &                          -0.1069 &  0 &  0.1479 &       0.1115 &   0  &  0.1135 \\
 $\sineff{e}$ &                   1.413 &  0 &  -1.821 &        -1.426 &   0  &  -1.423 \\
 $\sineff{b}$ &                  1.415 &  0 &  -1.821 &        -1.427 &   0  &  -1.417 \\
 $\sineff{c}$ &                  1.413 &  0 &  -1.821 &        -1.426 &   0  &  -1.423 \\
 $\Asy{e}$ &                   -17.61 &  0 &  22.71 &         17.78 &   0  &  17.74 \\
 $\Asy{b}$ &                 -0.2268 &  0 &  0.2876 &       0.2287 &   0  &  0.2271 \\
 $\Asy{c}$ &                    -1.697 &  0 &  2.192 &         1.713 &   0  &  1.710 \\
 $\AFB{e}$ &                    -35.22 &  0 &  45.41 &         35.56 &   0  &  35.48 \\
 $\AFB{b}$ &                   -17.84 &  0 &  22.99 &         18.01 &   0  &  17.97 \\
 $\AFB{c}$ &                    -19.31 &  0 &  24.90 &         19.50 &   0  &  19.45 \\
    \hline
  \end{tabular}
\end{center}
\caption{The $b$ coefficients defined in \eq{bdef}, characterizing the shift in the output observables due to new physics that shifts vector boson self-energies.}\label{table:selfenergies}
\end{table}
\begin{itemize}
\item $b_{i,zz}^{\prime}$, $b_{i,ww}$ are the simplest, since they vanish for most of the observables. In particular, $b_{i,zz}^{\prime}$, which comes from wavefunction renormalization, is nonzero only for $Z$ boson decay widths:
    \be
    b_{\Gamma_f,zz}^{\prime} = b_{\Gamma\inv,zz}^{\prime} = b_{\Gamma\had,zz}^{\prime} = b_{\Gamma_Z,zz}^{\prime} = 1.
    \ee
    Note that wavefunction renormalization cancels out in $\sigma\had$, and ratios of decay widths. $b_{i,ww}$ is related to the shift in the $W$ boson mass, so is nonzero only for:
    \be
    b_{m_W,ww} = \frac{1}{2}.
    \ee
\item $b_{i,zz}$, $b_{i,\gamma\gamma}^{\prime}$, $b_{i,ww}^0$ are simply related to $d_{i,m_Z}$, $d_{i,\alpha}$, $d_{i,G_F}$, respectively. Since $\piggp$, $\piwwo$ only enter $\almz$, $G_F$, respectively, we have
    \be
    b_{i,\gamma\gamma}^{\prime} = -d_{i,\alpha},\quad b_{i,ww}^0 = d_{i,G_F}
    \ee
    for all $\Obs{i}{}$. Similarly,
    \be
    b_{i,zz} = -\frac{1}{2}d_{i,m_Z}
    \ee
    except for those observables having direct dependence on the $Z$ boson mass:
    \bea
    b_{i,zz} &=& \frac{1}{2} (1-d_{i,m_Z})\quad\text{for } i=\Gamma_f, \Gamma\inv, \Gamma\had, \Gamma_Z,\\
    b_{\sigma\had,zz} &=& -\frac{1}{2} (2+d_{i,m_Z}).
    \eea
\item Finally, $b_{i,\gamma z}$ should be derived from the dependence on $\sineff{f}$. For the $Z$ partial widths, it can be read off from \eq{Gammafpi}:
\be
b_{\Gamma_f,\gamma z} = a_f,\quad b_{\Gamma\inv,\gamma z} = 3a_{\nu} = 0,
\ee
with $a_f$ given in \eqs{afana}{afnum}. For $i=\Gamma\had,\Gamma_Z$, $b_{i,\gamma z}$ is a weighted sum. At leading order:
\bea
b_{\Gamma\had,\gamma z} &=& \sum_{f\in\text{had}}\frac{\Gamma_f}{\Gamma\had}b_{\Gamma_f,\gamma z} = \frac{\sum_{f\in\text{had}}\bigl[1+(1-4|Q_f|s^2)^2\bigr]b_{\Gamma_f,\gamma z}}{\sum_{f\in\text{had}}\bigl[1+(1-4|Q_f|s^2)^2\bigr]},\\
b_{\Gamma_Z,\gamma z} &=& \sum_f\frac{\Gamma_f}{\Gamma_Z}b_{\Gamma_f,\gamma z} = \frac{\sum_f\bigl[1+(1-4|Q_f|s^2)^2\bigr]b_{\Gamma_f,\gamma z}}{\sum_f\bigl[1+(1-4|Q_f|s^2)^2\bigr]}.
\eea
For the ratios of partial widths, and the $Z$-pole cross section:
\bea
b_{R_\ell,\gamma z} = b_{\Gamma\had,\gamma z}-b_{\Gamma_\ell,\gamma z},\quad b_{R_q,\gamma z} = b_{\Gamma_q,\gamma z}-b_{\Gamma\had,\gamma z},\quad
b_{\sigma\had,\gamma z} = b_{\Gamma_e,\gamma z}+b_{\Gamma\had,\gamma z}-2b_{\Gamma_Z,\gamma z}.
\eea
For the asymmetry observables, we can read off from \eq{seffpi}:
\be
b_{\sineff{f},\gamma z} = -\frac{c}{s}.
\ee
And hence, at leading order,
\bea
b_{\Asy{f},\gamma z} &=& \frac{s^2}{\tree{\Asy{f}}}\frac{\partial\tree{\Asy{f}}}{\partial(s^2)}b_{\sineff{f},\gamma z} = \frac{4|Q_f|sc[1-(1-4|Q_f|s^2)^2]}{(1-4|Q_f|s^2)[1-(1+4|Q_f|s^2)^2]},\\
b_{\AFB{f},\gamma z} &=& b_{\Asy{e},\gamma z} + b_{\Asy{f},\gamma z}.
\eea
\end{itemize}
The numerical values for these $b$ coefficients are listed in Table~\ref{table:selfenergies}. The calculation is done with $s^2=0.231620$, and the sign conventions for the gauge couplings are $g_1>0$, $g_2>0$ (hence $s>0$).

\section{Conclusion}\label{sec:summary}

In this paper we presented an expansion formalism that facilitates precision electroweak analysis. By recasting all observables in terms of six very well measured input observables, we can calculate each of them easily by expanding about the reference values of the input observables, chosen in accord with experimental measurements. Also, the  formalism developed here can be applied in a simple manner to calculate new physics corrections to electroweak observables and derive constraints on new physics models. Some examples were worked out for illustration.

For numerical results we calculated the reference values and expansion coefficients using the ZFITTER package. Most, though not all, of these results reflect state-of-the-art calculations in the literature. Various higher order calculations of electroweak observables have been done since the release of ZFITTER 6.42 in 2005, but their impact on precision analysis is not significant at present because the power of the precision program is limited by experimental errors. However, improvements of our results to better accuracy with the inclusion of these and future calculations may be necessary in the future, if  experimental priorities of next-generation facilities involve Giga-Z or Tera-Z options~\cite{Heinemeyer:2010yp,Gomez-Ceballos:2013zzn}. With $10^9$ or $10^{12}$ $Z$ bosons produced at a future collider, unprecedented levels of reliable theoretical calculations will be needed to meet the unprecedented levels of experimental accuracy. We hope that the formalism presented here, with improving numerical results, will continue to be helpful for efficient and reliable calculations of SM results and beyond the SM corrections in the precision electroweak program.

\bigskip
\noindent
{\bf Acknowledgments:} This work was supported in part by the Department of Energy. We wish to thank T.~Riemann for helpful communications regarding ZFITTER, and A.~Freitas for pointing out a mistake in our implementation of higher-order QCD corrections in an earlier version.

\appendix
\section{Technical details of ZFITTER}\label{app:zfitter}

We rely on ZFITTER 6.42 for all numerical calculations of observables, and obtain the expansion coefficients $c_{ii'}$, $c_{ii'j'}$ by numerical differentiation. Some calculational details are presented in this appendix.

We use the DIZET package in ZFITTER, modified slightly to allow for $G_F$ as input. The flags are set to default listed in~\cite{Arbuzov:2005ma}, with the following exceptions:
\begin{itemize}
  \item \flag{NPAR(7)} = \flag{IALEM} = 2 (default = 3) to allow for $\dalhad$ as input.
  \item \flag{NPAR(20)} = \flag{IGFER} =3 (default = 2) to allow for $G_F$ as input. Note that the only available options for this flag in ZFITTER are 0, 1, 2, and none of them allows treats $G_F$ as input (since it is extraordinarily well measured), but we added a new option 3 to be consistent with the modification of the codes mentioned above.
\end{itemize}
In principle, alternative choices for the flags are possible. But to be consistent with our formalism, the following flags should not be changed from default:
\begin{itemize}
  \item \flag{NPAR(2)} = \flag{IAMT4} (default = 4): 4 is the only option consistent with treating $G_F$ as input.
  \item \flag{NPAR(4)} = \flag{IMOMS} (default = 1): 1 treats $m_Z$ as input and $m_W$ as output, not otherwise.
\end{itemize}

The derivatives appearing in $c_{ii'}$ [\eq{ciip}] are carried out numerically via~\cite{Press:recipes}
\beq{numdiff}
\frac{\partial\Obs{i}{SM}}{\partial\Obs{i'}{SM}} \simeq \frac{\Obs{i}{SM}\bigr|_{(1+h)\Obs{i'}{ref}}-\Obs{i}{SM}\bigr|_{(1-h)\Obs{i'}{ref}}}{2h\Obs{i'}{ref}},
\eeq
where $h$ is chosen differently for different input observables; see Table~\ref{table:h}. The choices are made empirically, and are expected to be optimal in reducing the combination of truncation and roundoff errors.\footnote{We calculated the derivatives with $h$ varied within a wide range, and recognized the regime where the results fluctuate (roundoff error dominates) and the regime where the results vary monotonically (truncation error dominates). The optimal $h$ is in between these two regimes. In principle, the optimal $h$ can be determined from the machine precision and the algorithm for evaluating the functions. But in practice, this is difficult due to the complexity of calculations in ZFITTER, so we took this empirical approach.} We found that the numerical errors typically occur at the 7th or 8th digit, and thus do not affect the digits presented in the tables earlier in this paper.

For calculating $c_{ii'j'}$ [\eq{ciipjp}], on the other hand, we make use of the fact that
\be
c_{ii'j'} = \left[\Obs{j'}{SM}\frac{\partial c_{ii'}}{\partial\Obs{j'}{SM}} + c_{ii'} c_{ij'} - \delta_{i'j'} c_{ii'}\right]\Biggr|_{\Obs{i'}{}=\Obs{i'}{ref}}.
\ee
and evaluate the derivatives with the same $h$ mentioned above.

\begin{table}
\begin{center}
  \begin{tabular}{|c|c|c|c|c|c|c|}
    \hline
    $\Obs{i'}{}$ & $m_Z$ & $G_F$ & $\dalhad$ & $m_t$ & $\alsmz$ & $m_H$ \\
    \hline
    $h$ & $10^{-6}$ & $10^{-5}$ & $10^{-4}$ & $10^{-4}$ & $10^{-4}$ & $10^{-4}$ \\
    \hline
  \end{tabular}
\end{center}
\caption{The $h$ chosen for each input observable in numerical differentiation. See \eq{numdiff}}\label{table:h}
\end{table}

\section{QCD corrections to $Z$ decay}\label{app:qcd}

In this appendix we discuss the calculation of $\Gamma_q$. As was mentioned in Section~\ref{sec:beyond}, this discussion is motivated by two features in our numerical results. First, the uncertainty in $c_{R_b,\alpha_s}$ is much larger than that in all other expansion coefficients. Second, $c_{\Gamma_q,\alpha_s}$, which characterize the sensitivity of $Z\to q\bar q$ partial widths to the strong coupling constant, are very different for different quarks (see Table~\ref{table:correct}), though at leading order QCD corrections are flavor-universal. This second feature led us to investigate and confirm the reliability of our numerical calculation. Both features are related to $\Ord{\alpha_s^2}$ corrections, as we will explain in the following.

\begin{table}
\begin{center}
  \begin{tabular}{|c|c|c|c|c|c|c|}
    \hline
    $q$ & $u$ & $c$ & $d$ & $s$ & $b$ & had \\
    \hline
    $c_{\Gamma_q,\alpha_s}$ & 0.04892 & 0.05046 & 0.02697 & 0.02697 & 0.03672 & 0.03690 \\
    \hline
  \end{tabular}
\end{center}
\caption{Numerical values of $c_{\Gamma_q,\alpha_s}$ and $c_{\Gamma\had,\alpha_s}$. The difference among these numbers is explained in the text. Note that $\Gamma_{u,d,s}$ are not in our observables list, since they are practically unmeasurable.}\label{table:correct}
\end{table}

Following the notations in ZFITTER~\cite{Bardin:1999yd}, we write the formula that calculates the partial width of the Z boson to $q\bar q$ as follows:
\beq{Gammaq}
\Gamma_q = 3 \Gamma_0 \left|\rho_Z^q\right| \left(\left|g_Z^q\right|^2 R_V^q + R_A^q\right) + \Delta_{\text{EW/QCD}},
\eeq
where
\be
\Gamma_0 = \frac{G_F m_Z^3}{24\sqrt 2 \pi} \simeq 83\text{ MeV}.
\ee
$\rho_Z^q$ and $g_Z^q$ are effective couplings that incorporate electroweak loop corrections to the $Z$ decay; in particular, $g_Z^q$ is the ratio of effective vector and axial couplings. $R_V^q$ and $R_A^q$ are vector and axial radiator functions, which deal with final state QCD and QED radiation. There is also an additive mixed EW/QCD correction term $\Delta_{\text{EW/QCD}}$ that does not factorize.

The radiator functions $R_V^q$ and $R_A^q$ actually depend on the energy scale. In \eq{Gammaq} it is implicit that they are evaluated at the $Z$ mass. Explicitly, the vector radiator function is given by
\beq{RV}
R_V^q = 1 + \frac{3}{4} Q_q^2 \frac{\alpha}{\pi} + \frac{\alpha_s}{\pi} - \frac{1}{4} Q_q^2 \frac{\alpha}{\pi} \frac{\alpha_s}{\pi} + \biggl[C_{02} + C_2^t\Bigl(\frac{m_Z^2}{m_t^2}\Bigr)\biggr] \Bigl(\frac{\alpha_s}{\pi}\Bigr)^2 + C_{03} \Bigl(\frac{\alpha_s}{\pi}\Bigr)^3 + \Ord{\alpha^2}, \Ord{\alpha_s^4}, \Ord{m_q^2},
\eeq
where
\bea
C_{02} &=& \frac{365}{24} - 11\zeta(3) + \Bigl[-\frac{11}{12} + \frac{2}{3}\zeta(3)\Bigr] n_q,\\
C_2^t(x) &=& x\Bigl(\frac{44}{675} - \frac{2}{135} \ln x\Bigr) + \Ord{x^2},\\
C_{03} &=&\frac{87029}{288} - \frac{121}{8}\zeta(2) - \frac{1103}{4}\zeta(3) + \frac{275}{6}\zeta(5)\nonumber\\
 &\quad& + \Bigl[-\frac{7847}{216} + \frac{11}{6}\zeta(2) + \frac{262}{9}\zeta(3) -\frac{25}{9}\zeta(5)\Bigr] n_q \nonumber\\
&\quad& + \Bigl[\frac{151}{162} - \frac{1}{18}\zeta(2) - \frac{19}{27}\zeta(3)\Bigr] n_q^2.
\eea
$\zeta$ is the Riemann zeta function. At the $Z$ pole the number of light quark flavors $n_q=5$.

To the order shown in \eq{RV}, $R_A^q$ receives additional contributions at $\Ord{\alpha_s^2}$ and $\Ord{\alpha_s^3}$:
\be
R_A^q = R_V^q - 2T^{(3)}_q\left[ I^{(2)} \biggl(\frac{m_Z^2}{m_t^2}\biggr) \Bigl(\frac{\alpha_s}{\pi}\Bigr)^2 + I^{(3)} \biggl(\frac{m_Z^2}{m_t^2}\biggr) \Bigl(\frac{\alpha_s}{\pi}\Bigr)^3\right] + \Ord{\alpha^2}, \Ord{\alpha_s^4}, \Ord{m_q^2},
\ee
where $T^{(3)}_q=+\frac{1}{2}$ ($-\frac{1}{2}$) for up (down) type quarks, and
\bea
I^{(2)}(x) &=& -\frac{37}{12} + \ln x + \frac{7}{81} x + \frac{79}{6000} x^2 + \Ord{x^3},\\
I^{(3)}(x) &=& -\frac{5075}{216} + \frac{23}{6}\zeta(2) + \zeta(3) + \frac{67}{18}\ln x + \frac{23}{12} \ln^2 x + \Ord{x}.
\eea
These terms are called singlet axial corrections. $I^{(2)}$ was first calculated in~\cite{Kniehl:1989bb,Kniehl:1989qu}. There the focus was on the total hadronic width, and the singlet axial corrections (approximately) cancel among the ``light'' quarks $u,d,c,s$. However, these terms are visible in each partial width, and are numerically comparable to the $\Ord{\alpha_s}$ terms. Being negative, they make $c_{\Gamma_u,\alpha_s}$, $c_{\Gamma_c,\alpha_s}$ larger than $c_{\Gamma_d,\alpha_s}$, $c_{\Gamma_s,\alpha_s}$.

We might expect $c_{\Gamma_b,\alpha_s}$ to be close to $c_{\Gamma_d,\alpha_s}$, $c_{\Gamma_s,\alpha_s}$, but in Table~\ref{table:correct} it is seen to be larger. This is due to a positive contribution from the $\Ord{m_q^2}$ terms, which are significant only for the $b$ quark. To be precise, $m_q$ in these terms should be taken as the running masses at the $Z$ pole, obtained by solving RG equations. For the $b$ quark, the dependence of these RG equations on $\alpha_s$ is strong enough to overcome the $\frac{m_b^2}{m_Z^2}$ suppression, and the contribution to $c_{\Gamma_b,\alpha_s}$ turns out to be positive. Similarly, $c_{\Gamma_c,\alpha_s}$ also receives a positive contribution, which explains the small difference from $c_{\Gamma_u,\alpha_s}$.

Now that we have understood the difference among $c_{\Gamma_q,\alpha_s}$ and are confident about their numerical values, we can calculate $c_{\Gamma\had,\alpha_s}$ by a weighted average, and the result is, by accident, very close to $c_{\Gamma_b,\alpha_s}$ (see Table~\ref{table:correct}). As a result, $c_{R_b,\alpha_s}=c_{\Gamma_b,\alpha_s}-c_{\Gamma\had,\alpha_s}$ is much smaller than either of $c_{\Gamma_b,\alpha_s}$, $c_{\Gamma\had,\alpha_s}$, and can thus have large uncertainty though the uncertainties in the latter are small.

Finally, a few comments are in order regarding future improvements of the $Z$ decay calculation. Recent developments, including the complete $\Ord{\alpha_s^4}$ QCD corrections~\cite{Baikov:2012er,Baikov:2012xh} and fermionic electroweak two-loop corrections~\cite{Freitas:2014hra} will be implemented in future versions of ZFITTER~\cite{Akhundov:2014era}, which will certainly help improve the accuracy of our results. Meanwhile, we note two other aspects of the ZFITTER calculation that could be improved. First, the $\Delta_{\text{EW/QCD}}$ term in \eq{Gammaq} is implemented as fixed numbers in ZFITTER, so the dependence on input observables is lost, which is especially relevant in the expansion formalism. Second, the $\Ord{\alpha_s^3}$ difference between $\Gamma\had$ and $\sum_q\Gamma_q$ mentioned in a footnote in Section~\ref{sec:SMpO}, though calculated and stored in \flag{ZPAR(29)}=\flag{QCDCOR(13)}, is not included in the calculation of $\Gamma\had$ or the total width $\Gamma_Z$. The size of this term is only on the order of $10^{-5} \,\Gamma\had$~\cite{Chetyrkin:1996ia}, but the error might be magnified when the expansion coefficients are calculated.



\begin{thebibliography}{99}

 \bibitem{Aad:2012tfa}
  G.~Aad {\it et al.}  [ATLAS Collaboration],
  ``Observation of a new particle in the search for the Standard Model Higgs boson with the ATLAS detector at the LHC,''
  Phys.\ Lett.\ B {\bf 716}, 1 (2012)
  [arXiv:1207.7214 [hep-ex]].

\bibitem{Chatrchyan:2012ufa}
  S.~Chatrchyan {\it et al.}  [CMS Collaboration],
  ``Observation of a new boson at a mass of 125 GeV with the CMS experiment at the LHC,''
  Phys.\ Lett.\ B {\bf 716}, 30 (2012)
  [arXiv:1207.7235 [hep-ex]].

\bibitem{Chatrchyan:2012jja}
  S.~Chatrchyan {\it et al.}  [CMS Collaboration],
  ``Study of the Mass and Spin-Parity of the Higgs Boson Candidate Via Its Decays to Z Boson Pairs,''
  Phys.\ Rev.\ Lett.\  {\bf 110}, 081803 (2013)
  [arXiv:1212.6639 [hep-ex]].

\bibitem{Almeida:2013jfa}
  L.~G.~Almeida, S.~J.~Lee, S.~Pokorski and J.~D.~Wells,
  ``Study of the 125 GeV Standard Model Higgs Boson Partial Widths and Branching Fractions,''
  Phys.\ Rev.\ D {\bf 89}, 033006 (2014)
  [arXiv:1311.6721 [hep-ph]].

\bibitem{Bardin:1999yd}
  D.~Y.~.Bardin, P.~Christova, M.~Jack, L.~Kalinovskaya, A.~Olchevski, S.~Riemann and T.~Riemann,
  ``ZFITTER v.6.21: A Semianalytical program for fermion pair production in $e^+ e^-$ annihilation,''
  Comput.\ Phys.\ Commun.\  {\bf 133}, 229 (2001)
  [hep-ph/9908433].

\bibitem{Arbuzov:2005ma}
  A.~B.~Arbuzov, M.~Awramik, M.~Czakon, A.~Freitas, M.~W.~Grunewald, K.~Monig, S.~Riemann and T.~Riemann,
  ``ZFITTER: A Semi-analytical program for fermion pair production in $e^+ e^-$ annihilation, from version 6.21 to version 6.42,''
  Comput.\ Phys.\ Commun.\  {\bf 174}, 728 (2006)
  [hep-ph/0507146].

\bibitem{Beringer:1900zz}
  J.~Beringer {\it et al.}  [Particle Data Group Collaboration],
  ``Review of Particle Physics (RPP),''
  Phys.\ Rev.\ D {\bf 86}, 010001 (2012).
  And 2013 partial update for the 2014 edition.

  \bibitem{constants}
  See \url{http://physics.nist.gov/cuu/Constants/}

\bibitem{Chetyrkin:1996ia}
  K.~G.~Chetyrkin, J.~H.~Kuhn and A.~Kwiatkowski,
  ``QCD corrections to the $e^{+} e^{-}$ cross-section and the $Z$ boson decay rate: Concepts and results,''
  Phys.\ Rept.\  {\bf 277}, 189 (1996).

\bibitem{ALEPH:2005ab}
  S.~Schael {\it et al.}  [ALEPH and DELPHI and L3 and OPAL and SLD and LEP Electroweak Working Group and SLD Electroweak Group and SLD Heavy Flavour Group Collaborations],
  ``Precision electroweak measurements on the $Z$ resonance,''
  Phys.\ Rept.\  {\bf 427}, 257 (2006)
  [hep-ex/0509008].

\bibitem{Abazov:2011ws}
  V.~M.~Abazov {\it et al.}  [D0 Collaboration],
  ``Measurement of $\sin^2\theta_{\rm eff}^{\ell}$ and $Z$-light quark couplings using the forward-backward charge asymmetry in $p\bar{p} \to Z/\gamma^{*} \to e^{+}e^{-}$ events with ${\cal L}=5.0$ fb$^{-1}$ at $\sqrt{s}=1.96$ TeV,''
  Phys.\ Rev.\ D {\bf 84}, 012007 (2011)
  [arXiv:1104.4590 [hep-ex]].

\bibitem{Han:2011vw}
  J.~Han [CDF Collaboration],
  ``The Angular Coefficients and $A_{fb}$ of Drell-Yan $e^{+}e^{-}$ Pairs in the Z Mass Region from $p\bar{p}$ Collision at $\sqrt{s}$ = 1.96 TeV,''
  arXiv:1110.0153 [hep-ex].

\bibitem{Group:2012gb}
  Electroweak Working Group [CDF and D0 Collaborations],
  ``2012 Update of the Combination of CDF and D0 Results for the Mass of the W Boson,''
  arXiv:1204.0042 [hep-ex].

\bibitem{Alcaraz:2006mx}
  J.~Alcaraz {\it et al.}  [ALEPH and DELPHI and L3 and OPAL and LEP Electroweak Working Group Collaborations],
  ``A Combination of preliminary electroweak measurements and constraints on the standard model,''
  hep-ex/0612034.

\bibitem{CDF:2013jga}
  M.~Muether {\it et al.}  [Tevatron Electroweak Working Group and CDF and D0 Collaborations],
  ``Combination of CDF and DO results on the mass of the top quark using up to 8.7 ${\rm fb}^{-1}$ at the Tevatron,''
  arXiv:1305.3929 [hep-ex].

\bibitem{Buchmuller:1985jz}
  W.~Buchmuller and D.~Wyler,
  ``Effective Lagrangian Analysis of New Interactions and Flavor Conservation,''
  Nucl.\ Phys.\ B {\bf 268}, 621 (1986).

\bibitem{Grzadkowski:2010es}
  B.~Grzadkowski, M.~Iskrzynski, M.~Misiak and J.~Rosiek,
  ``Dimension-Six Terms in the Standard Model Lagrangian,''
  JHEP {\bf 1010}, 085 (2010)
  [arXiv:1008.4884 [hep-ph]].

\bibitem{Peskin:2001rw}
  M.~E.~Peskin and J.~D.~Wells,
  ``How can a heavy Higgs boson be consistent with the precision electroweak measurements?,''
  Phys.\ Rev.\ D {\bf 64}, 093003 (2001)
  [hep-ph/0101342].

\bibitem{Bamert:1996px}
See, e.g., P.~Bamert, C.~P.~Burgess, J.~M.~Cline, D.~London and E.~Nardi,
``$R_b$ and new physics: A Comprehensive analysis,''
  Phys.\ Rev.\ D {\bf 54}, 4275 (1996)
  [hep-ph/9602438].

\bibitem{Peskin:1991sw}
 An early comprehensive paper on this possibility is M.~E.~Peskin and T.~Takeuchi,
  Phys.\ Rev.\ D {\bf 46}, 381 (1992).

\bibitem{Wells:2005vk}
  J.~D.~Wells,
  ``TASI lecture notes: Introduction to precision electroweak analysis,''
  hep-ph/0512342.

\bibitem{Heinemeyer:2010yp}
  S.~Heinemeyer and G.~Weiglein,
  ``Top, GigaZ, MegaW,''
  arXiv:1007.5232 [hep-ph].

\bibitem{Gomez-Ceballos:2013zzn}
  M.~Bicer {\it et al.}  [TLEP Design Study Working Group Collaboration],
  ``First Look at the Physics Case of TLEP,''
  JHEP {\bf 1401}, 164 (2014)
  [arXiv:1308.6176 [hep-ex]].

  \bibitem{Press:recipes}
  W.H.~Press, S.A.~Teukosky, W.T.~Vetterling, B.P.~Flannery,
  {\it Numerical Recipes in C: The Art of Scientific Computing}, 2nd ed.,
  Cambridge University Press, 1992.

\bibitem{Kniehl:1989bb}
  B.~A.~Kniehl and J.~H.~Kuhn,
  ``QCD Corrections to the Axial Part of the Z Decay Rate,''
  Phys.\ Lett.\ B {\bf 224}, 229 (1989).

\bibitem{Kniehl:1989qu}
  B.~A.~Kniehl and J.~H.~Kuhn,
  ``QCD Corrections to the Z Decay Rate,''
  Nucl.\ Phys.\ B {\bf 329}, 547 (1990).

\bibitem{Baikov:2012er}
  P.~A.~Baikov, K.~G.~Chetyrkin, J.~H.~Kuhn and J.~Rittinger,
  ``Complete ${\cal O}(\alpha_s^4)$ QCD Corrections to Hadronic $Z$-Decays,''
  Phys.\ Rev.\ Lett.\  {\bf 108}, 222003 (2012)
  [arXiv:1201.5804 [hep-ph]].

\bibitem{Baikov:2012xh}
  P.~A.~Baikov, K.~G.~Chetyrkin, J.~H.~Kuhn and J.~Rittinger,
  ``$R(s)$ and $Z$ decay in order $\alpha_s^4$: complete results,''
  PoS RADCOR2011, 030 (2011)
  [PoS RADCOR {\bf 2011}, 030 (2011)]
  [arXiv:1210.3594 [hep-ph]].

\bibitem{Freitas:2014hra}
  A.~Freitas,
  ``Higher-order electroweak corrections to the partial widths and branching ratios of the $Z$ boson,''
  JHEP {\bf 1404}, 070 (2014)
  [arXiv:1401.2447 [hep-ph]].

\bibitem{Akhundov:2014era}
  A.~Akhundov, A.~Arbuzov, S.~Riemann and T.~Riemann,
  ``The ZFITTER project,''
  Phys.\ Part.\ Nucl.\  {\bf 45}, no. 3, 529 (2014)
  [arXiv:1302.1395 [hep-ph]].
  And private communication with T.~Riemann.

\end{thebibliography}
\end{document}